\documentclass[12pt,A4]{article}
\usepackage{psfrag}
\usepackage{graphics,graphicx}
\usepackage{amssymb,epsfig,amsmath,euscript,array}
\usepackage[nosort]{cite}
\usepackage{pstricks}
\usepackage{color}
\usepackage{tikz}
\usepackage{float}
\usepackage{caption}
\usepackage{subcaption}
\usepackage{caption}
\usepackage[T1]{fontenc}
\usepackage[unicode=true,pdfusetitle,
  bookmarks=true,bookmarksnumbered=true,bookmarksopen=true,
breaklinks=true,pdfborder={0 0 0},backref=false,colorlinks=true]{hyperref}

\hypersetup{ linkcolor=black, citecolor=black, urlcolor=black}

\makeatletter
\@addtoreset{equation}{section}
\makeatother

\newcounter{multieqs}

\newcommand{\be}{\begin{equation}}
\newcommand{\ee}{\end{equation}}

\def\lt{\tilde{\l}}

\newcommand{\bm}[1]{\mbox{\boldmath $#1$}}

\newcommand{\kslash}{k \!\!\! / }

\newcommand{\lslash}{l \!\! / }
\newcommand{\Pslash}{P \!\!\!\! / }

\newcommand{\islash}{i \!\!\! / }
\newcommand{\jslash}{j \!\!\! / }
\newcommand{\aslash}{a \!\!\! / }
\newcommand{\bslash}{{b \hspace{-6pt} \slash} }

\newcommand{\onslash}{1 \!\!\! / }
\newcommand{\twslash}{2 \!\!\!/ }
\newcommand{\thslash}{3 \!\!\!/ }
\newcommand{\foslash}{4 \!\!\! / }
\newcommand{\fislash}{5 \!\!\! / }

\newcommand{\mslash}{m \!\!\! / }

\def\bd{\begin{document}}
\def\ed{\end{document}}
\def\nn{\nonumber}
\def\bea{\begin{eqnarray}}
\def\eea{\end{eqnarray}}

\def\ab{(ijab)}
\def\ba{(ijba)}
\def\ijab{{\tr}_{+}(\islash\, \jslash\, \aslash \, \bslash)}
\def\ijba{{\tr}_{+}(\islash\, \jslash\, \bslash \, \aslash)}
\def\ijaP{{\tr}_{+}(\islash\, \jslash\, \aslash \, \Pslash)}
\def\ijPLa{{\tr}_{+}(\islash\, \jslash\, \Pslash_L \, \aslash)}
\def\ijaPL{{\tr}_{+}(\islash\, \jslash\, \aslash \, \Pslash_L)}
\def\ijPLza{{\tr}_{+}(\islash\, \jslash\, \Pslash_{L;z} \, \aslash)}
\def\ijaPLz{{\tr}_{+}(\islash\, \jslash\, \aslash \, \Pslash_{L;z})}
\def\ijPa{{\tr}_{+}(\islash\, \jslash\, \Pslash \, \aslash)}
\def\iaPb{{\tr}_{+}(\islash\, \aslash\, \Pslash \, \bslash)}
\def\ibPa{{\tr}_{+}(\islash\, \bslash\, \Pslash \, \aslash)}
\def\ijPmu{{\tr}_{+}(\islash\, \jslash\, \Pslash \, \mu)}
\def\ibmuP{{\tr}_{+}(\islash\, \bslash\, \mu \, \Pslash)}
\def\ibmua{{\tr}_{+}(\islash\, \bslash\, \mu \, \aslash)}
\def\iamub{{\tr}_{+}(\islash\, \aslash\, \mu \, \bslash)}
\def\jaPb{{\tr}_{+}(\jslash\, \aslash\, \Pslash \, \bslash)}
\def\ijmuP{{\tr}_{+}(\islash\, \jslash\, \mu \, \Pslash)}
\def\ijmum{{\tr}_{+}(\islash\, \jslash\, \mu \, \mslash)}
\def\ijmmu{{\tr}_{+}(\islash\, \jslash\, \mslash \, \mu)}
\def\ijmP{{\tr}_{+}(\islash\, \jslash\, \mslash \, \Pslash)}
\def\iabP{{\tr}_{+}(\islash\, \aslash\, \bslash \, \Pslash)}
\def\ijbP{{\tr}_{+}(\islash\, \jslash\, \bslash \, \Pslash)}
\def\jbPa{{\tr}_{+}(\jslash\, \bslash\, \Pslash \, \aslash)}
\def\ijPb{{\tr}_{+}(\islash\, \jslash\, \Pslash \, \bslash)}
\def\jbmua{{\tr}_{+}(\jslash\, \bslash\, \mu \, \aslash)}

\def\loablt{ {\tr}_{+}(\lslash_1\, \aslash \, \bslash\, \lslash_2)}

\def\ijlolt{{\tr}_{+}(\islash\, \jslash\, \lslash_1 \, \lslash_2)}
\def\ijltlo{{\tr}_{+}(\islash\, \jslash\, \lslash_2 \, \lslash_1)}
\def\ibloa{{\tr}_{+}(\islash\, \bslash\, \lslash_1 \, \aslash)}
\def\jaltb{{\tr}_{+}(\jslash\, \aslash\, \lslash_2 \, \bslash)}
\def\ialtb{{\tr}_{+}(\islash\, \aslash\, \lslash_2 \, \bslash)}
\def\bltloa{{\tr}_{+}(\bslash\, \lslash_2\, \lslash_1 \, \aslash)}
\def\jbloa{{\tr}_{+}(\jslash\, \bslash\, \lslash_1 \, \aslash)}
\def\ibPb{{\tr}_{+}(\islash\, \bslash\, \Pslash \, \bslash)}
\def\ijltb{{\tr}_{+}(\islash\, \jslash\, \lslash_2 \, \bslash)}

\def\ijloa{{\tr}_{+}(\islash\, \jslash\,  \lslash_1 \, \aslash)}
\def\ijblt{{\tr}_{+}(\islash\, \jslash\,  \bslash \, \lslash_2)}

\def\jakb{{\tr}_{+}(\jslash\, \aslash\, \kslash \, \bslash)}
\def\iakb{{\tr}_{+}(\islash\, \aslash\, \kslash \, \bslash)}

\def\tofo{{\tr}_{+}(\onslash\, \thslash\, \twslash \, \foslash)}
\def\foto{{\tr}_{+}(\onslash\, \thslash\, \foslash \, \twslash)}
\def\tofi{{\tr}_{+}(\onslash\, \thslash\, \twslash \, \fislash)}
\def\fito{{\tr}_{+}(\onslash\, \thslash\, \fislash \, \twslash)}

\def\lrangle#1#2{\langle #1\,#2\rangle}

\def\Li{{$\rm Li}_2$}
\def\eps{\epsilon}
\def\epsuv{{\epsilon_{\rm \mbox{\tiny UV}}}}
\let\bm=\bibitem
\let\la=\label

\def\npb#1#2#3{Nucl. Phys. {\bf{B#1}} #3 (#2)}
\def\plb#1#2#3{Phys. Lett. {\bf{#1B}} #3 (#2)}
\def\prl#1#2#3{Phys. Rev. Lett. {\bf{#1}} #3 (#2)}
\def\prd#1#2#3{Phys. Rev. {D \bf{#1}} #3 (#2)}
\def\cmp#1#2#3{Comm. Math. Phys. {\bf{#1}} #3 (#2)}
\def\cqg#1#2#3{Class. Quantum Grav. {\bf{#1}} #3 (#2)}
\def\nppsa#1#2#3{Nucl. Phys. B (Proc. Suppl.) {\bf{#1A}}#3 (#2)}
\def\ap#1#2#3{Ann. of Phys. {\bf{#1}} #3 (#2)}
\def\ijmp#1#2#3{Int. J. Mod. Phys. {\bf{A#1}} #3 (#2)}
\def\rmp#1#2#3{Rev. Mod. Phys. {\bf{#1}} #3 (#2)}
\def\mpla#1#2#3{Mod. Phys. Lett. {\bf A#1} #3 (#2)}
\def\jhep#1#2#3{J. High Energy Phys. {\bf #1} #3 (#2)}
\def\atmp#1#2#3{Adv. Theor. Math. Phys. {\bf #1} #3 (#2)}
\newcommand{\EQ}[1]{\begin{equation} #1 \end{equation}}
\newcommand{\AL}[1]{\begin{subequations}\begin{align} #1 \end{align}\end{subequations}}
\newcommand{\SP}[1]{\begin{equation}\begin{split} #1 \end{split}\end{equation}}
\newcommand{\ALAT}[2]{\begin{subequations}\begin{alignat}{#1} #2 \end{alignat}
  \end{subequations}}
\def\beqa{\begin{eqnarray}}
\def\eeqa{\end{eqnarray}}
\def\beq{\begin{equation}}
\def\eeq{\end{equation}}
\def\sst{\scriptscriptstyle}
\def\thetabar{\bar\theta}
\def\Tr{{\rm Tr}}
\def\one{\mbox{1 \kern-.59em {\rm l}}}
 \def\Nh{\hat{N}}
\newcommand{\half}{{\textstyle {1 \over 2}}}

\def\a{\alpha}      \def\da{{\dot\alpha}}
\def\b{\beta}       \def\db{{\dot\beta}}
\def\c{\gamma}  \def\G{\Gamma}  \def\cdt{\dot\gamma}
\def\d{\delta}  \def\D{\Delta}  \def\ddt{\dot\delta}
\def\e{\epsilon}        \def\vare{\varepsilon}
\def\f{\phi}    \def\F{\Phi}    \def\vvf{\f}
\def\h{\eta}
\def\k{\kappa}
\def\l{\lambda} \def\L{\Lambda}
\def\m{\mu} \def\n{\nu}
\def\o{\omega}
\def\p{\pi} \def\P{\Pi}
\def\r{\rho}
\def\s{\sigma}  \def\S{\Sigma}
\def\t{\tau}
\def\th{\theta} \def\Th{\Theta} \def\vth{\vartheta}
\def\X{\Xeta}
\def\z{\zeta}
\def\de{\partial}
\def\cA{{\cal A}} \def\cB{{\cal B}} \def\cC{{\cal C}}
\def\cD{{\cal D}} \def\cE{{\cal E}} \def\cF{{\cal F}}
\def\cG{{\cal G}} \def\cH{{\cal H}} \def\cI{{\cal I}}
\def\cJ{{\cal J}} \def\cK{{\cal K}} \def\cL{{\cal L}}
\def\cM{{\cal M}} \def\cN{{\cal N}} \def\cO{{\cal O}}
\def\cP{{\cal P}} \def\cQ{{\cal Q}} \def\cR{{\cal R}}
\def\cS{{\cal S}} \def\cT{{\cal T}} \def\cU{{\cal U}}
\def\cV{{\cal V}} \def\cW{{\cal W}} \def\cX{{\cal X}}
\def\cY{{\cal Y}} \def\cZ{{\cal Z}}
\def\ua{\underline{\alpha}}
\def\ub{\underline{\phantom{\alpha}}\!\!\!\beta}
\def\uc{\underline{\phantom{\alpha}}\!\!\!\gamma}
\def\um{\underline{\mu}}
\def\ud{\underline\delta}
\def\ue{\underline\epsilon}
\def\una{\underline a}\def\unA{\underline A}
\def\unb{\underline b}\def\unB{\underline B}
\def\unc{\underline c}\def\unC{\underline C}
\def\und{\underline d}\def\unD{\underline D}
\def\une{\underline e}\def\unE{\underline E}
\def\unf{\underline{\phantom{e}}\!\!\!\! f}\def\unF{\underline F}
\def\unm{\underline m}\def\unM{\underline M}
\def\unn{\underline n}\def\unN{\underline N}
\def\unp{\underline{\phantom{a}}\!\!\! p}\def\unP{\underline P}
\def\unq{\underline{\phantom{a}}\!\!\! q}
\def\unQ{\underline{\phantom{A}}\!\!\!\! Q}
\def\unH{\underline{H}}
\def\As {{A \hspace{-6.4pt} \slash}\;}
\def\bs {{b \hspace{-6.4pt} \slash}\;}
\def\Ds {{D \hspace{-6.4pt} \slash}\;}
\def\ds {{\del \hspace{-6.4pt} \slash}\;}
\def\ss {{\s \hspace{-6.4pt} \slash}\;}
\def\ks {{ k \hspace{-6.4pt} \slash}\;}
\def\ps {{p \hspace{-6.4pt} \slash}\;}
\def\pas {{{p_1} \hspace{-6.4pt} \slash}\;}
\def\pbs {{{p_2} \hspace{-6.4pt} \slash}\;}
\def\Ps {{P \hspace{-6.4pt} \slash}\;}
\def\Qs {{Q \hspace{-6.4pt} \slash}\;}
\def\Fh{\hat{F}}
\def\Vh{\hat{V}}
\def\Xh{\hat{X}}
\def\ah{\hat{a}}
\def\xh{\hat{x}}
\def\yh{\hat{y}}
\def\ph{\hat{p}}
\def\xih{\hat{\xi}}
\def\psit{\tilde{\psi}}
\def\Psit{\tilde{\Psi}}
\def\tht{\tilde{\th}}
\def\lt{\tilde{\lambda}}
\def\hl{\hat{\lambda}}
\def\hlt{\hat{\tilde{\lambda}}}
\def\llt{\tilde{l}}
\def\At{\tilde{A}}
\def\Qt{\tilde{Q}}
\def\Rt{\tilde{R}}
\def\Nt{\tilde{N}}
\def\at{\tilde{a}}
\def\st{\tilde{s}}
\def\ft{\tilde{f}}
\def\pt{\tilde{p}}
\def\qt{\tilde{q}}
\def\vt{\tilde{v}}
\def\nt{\tilde{n}}
\def\delb{\bar{\partial}}
\def\bz{\bar{z}}
\def\bD{\bar{D}}
\def\bB{\bar{B}}
\def\bk{{\bf k}}
\def\bl{{\bf l}}
\def\bp{{\bf p}}
\def\bq{{\bf q}}
\def\br{{\bf r}}
\def\bx{{\bf x}}
\def\by{{\bf y}}
\def\bR{{\bf R}}
\def\bV{{\bf V}}
\def\d{\delta}\def\D{\Delta}\def\ddt{\dot\delta}
\def\pa{\partial} \def\del{\partial}
\def\xx{\times}
\def\uno{\mbox{1 \kern-.59em {\rm l}}}
\def\trp{^{\top}}
\def\inv{^{-1}}
\def\dag{{^{\dagger}}}
\def\pr{^{\prime}}
\def\lan{\langle}
\def\ran{\rangle}
\def\rar{\rightarrow}
\def\lar{\leftarrow}
\def\lrar{\leftrightarrow}
\newcommand{\0}{\,\!}      %this is just NOTHING!
\def\one{1\!\!1\,\,}
\def\im{\imath}
\def\jm{\jmath}
\newcommand{\tr}{\mbox{tr}}
\newcommand{\slsh}[1]{/ \!\!\!\! #1}
\def\vac{|0\rangle}
\def\lvac{\langle 0|}
\def\hlf{\frac{1}{2}}
\def\ove#1{\frac{1}{#1}}
\def\Box{\square}
\def\ZZ{\mathbb{Z}}
\def\CC#1{({\bf #1})}
\def\bcomment#1{}
\def\bfhat#1{{\bf \hat{#1}}}
\def\VEV#1{\left\langle #1\right\rangle}
\newcommand{\ex}[1]{{\rm e}^{#1}} \def\ii{{\rm i}}
\def\rr{{\rm r}} \def\rs{{\rm s}}\def\rv{{\rm v}}
\def\ri{{\rm i}}\def\rj{{\rm j}}
\newcommand{\lrbrk}[1]{\left(#1\right)}
\newcommand{\sfrac}[2]{{\textstyle\frac{#1}{#2}}}
\def\Li{{\rm Li}_2}
\DeclareMathOperator{\dif}{d \!}

\usetikzlibrary{trees}
\usetikzlibrary{calc}
\usetikzlibrary{decorations.pathmorphing}
\usetikzlibrary{decorations.markings}
\usetikzlibrary{intersections}
\usetikzlibrary{backgrounds}

\tikzset{
    photon/.style={decorate, decoration={snake}, draw=red},
    scalar/.style={thin, postaction={decorate}, decoration={markings,mark=at position .75 with {\arrow[draw=black]{>}}}},
    antiscalar/.style={thin, postaction={decorate}, decoration={markings,mark=at position .75 with {\arrow[draw=black]{<}}}},
    gluon/.style={thin,decorate, decoration={coil,amplitude=1.5pt, segment length=2pt}}
}

\newcommand{\fourbox}[5]{

  \begin{tikzpicture} [scale={#1}]
    \path[ use as bounding box] (-2,-3) rectangle (7,3);
    \tikzstyle{ellstyle}=[ fill=white,rounded corners=4pt,inner sep = 1pt, outer sep=1pt];

    \coordinate (aa) at ( $(3,0)+ (135:3)$);
    \coordinate (bb) at ( $(3,0)+ (45:3) $);
    \coordinate (cc) at ( $(3,0)+ (-45:3)$);
    \coordinate (dd) at ( $(3,0)+ (225:3)$);

    \draw
      [name path= propright
        [name path= propright,decoration={markings, mark=at position .5
          with {\arrow{latex}}}, postaction={decorate}]
    (aa) to (bb);
    \draw
     [name path= propdown  [name path= propdown,decoration={markings,
       mark=at position .5 with {\arrow{latex}}}, postaction={decorate}]
    (cc) to (bb);
    \draw
     [name path= propleft  [name path= propleft,decoration={markings,
       mark=at position .5 with {\arrow{latex}}}, postaction={decorate}]
    (dd) to (cc);
    \draw
     [name path= propup  [name path= propup,decoration={markings, mark=at
       position .5 with {\arrow{latex}}}, postaction={decorate}]
    (dd) to (aa);

    \draw
    ($(aa)$)-- ++(135:2);

    \draw
    ($(bb)$)--++(45:2);

    \draw
    ($(cc)$)--++(-45:2);

    \draw
    ($(dd)$)--++(225:2);

    \node[above]
    (l3) at ($(aa)!0.5!(bb)$) {#2};

    \node[right]
    (l4) at ($(bb)!0.5!(cc)$) {#3};

    \node[below]
    (l1) at  ($(cc)!0.5!(dd)$) {#4};

    \node[left]
    (l2) at  ($(dd)!0.5!(aa)$) {#5};

    \node[left]
    (ctopleft) at ($(aa)+(135:2)$) {$1$};

    \node[right]
    (ctopright) at ($(bb)+(45:2)$) {$2$};

    \node[right]
    (cbotright) at ($(cc)+(-45:2)$) {$3$};

     \node[left]
    (cbotleft) at ($(dd)+(225:2)$) {$4$};

  \end{tikzpicture}
}

\newcommand{\boxdiagram}{
  \begin{tikzpicture}[scale=0.5,cross/.style={path picture={
  \draw[black]
(path picture bounding box.south east) -- (path picture bounding box.north west) (path picture bounding box.south west) -- (path picture bounding box.north east);
}}]

    \draw[decoration={markings, mark=at position 0.5 with {\arrow{latex}}
    }, postaction={decorate}] (-2,2) --node[above] {$\ell_1$} (2,2);
    \draw (2,2) -- (2,-2);
    \draw (2,-2) -- (-2,-2);
    \draw (-2,-2) -- (-2,2);

    \draw[decoration={markings, mark=at position 0.5 with {\arrow{latex}}
    }, postaction={decorate}] (2,2) -- (3,3) node[above right] {$p_1$};
    \draw[decoration={markings, mark=at position 0.5 with {\arrow{latex}}
    }, postaction={decorate}] (2,-2) -- (3,-3) node[below right] {$p_2$};
    \draw[decoration={markings, mark=at position 0.5 with {\arrow{latex}}
    }, postaction={decorate}] (-2,-2) -- (-3,-3) node[below left] {$p_3$};
    \draw[decoration={markings, mark=at position 0.5 with {\arrow{latex}}
    }, postaction={decorate}] (-2,2) -- (-3,3) node[above left] {$p_4$};

  \end{tikzpicture}
}

\newcommand{\qqcutoneloop}{

  \begin{tikzpicture} [scale=0.5]
    \path[ use as bounding box] (-2,-3) rectangle (7,3);
    \tikzstyle{ellstyle}=[ fill=white,rounded corners=4pt,inner sep = 1pt, outer sep=1pt];
    \coordinate (fcentre) at (-1,0);
    \coordinate (acentre) at ($({7*cos(30)-sqrt(2)},0)$);
    \coordinate (aa) at ($(fcentre)+(45:1)$);
    \coordinate (ab) at ($(fcentre)+(-45:1)$);
    \coordinate (ba) at ($(acentre)+(135:1)$);
    \coordinate (bb) at ($(acentre)+(225:1)$);
    \coordinate (da) at ($(acentre)+(45:1)$);
    \coordinate (db) at ($(acentre)+(-45:1)$);

    \draw[name path= propright,decoration={markings, mark=at position .75 with {\arrow{latex}}}, postaction={decorate}] (aa) to [out = 45, in = 135, looseness = 1] (ba);

    \draw[name path= propleft,decoration={markings, mark=at position .75 with {\arrow{latex}}}, postaction={decorate}] (ab) to [out = -45, in = 225, looseness = 1] (bb);

    \draw[fill=gray!40] (fcentre) circle (1);
    \draw[fill=gray!40] (acentre) circle (1);

    \draw[name path = dashedcut,dashed] ($(aa)!0.5!(ba)+(0,1.5)$) -- ($(ab)!0.5!(bb)+(0,-1.5)$);

    \path [name intersections={of=propleft and dashedcut}] ;
    \coordinate (intleft) at (intersection-1);

    \path [name intersections={of=propright and dashedcut}] ;
    \coordinate (intright) at (intersection-1);

    \node(F) at (fcentre) {$F$};
    \node[above] (qq) at (-2.5,0) {$q$};

    \node[above right] (phA) at ($(da)+(45:1)$) {$\phi^{4}(p_2)$};
    \node[below right] (barph4) at ($(db)+(315:1)$) {$\bar\phi_{A}(p_1)$};

    \node[above right, ellstyle] (ellbir) at (intright) {$\ell_1$};
    \node[below right, ellstyle] (elliki) at (intleft) {$\ell_2$};

    \draw[decoration={markings, mark=at position .75 with {\arrow{latex}} }, postaction={decorate}] ($(da)$) -- ++(45:1);
    \draw[decoration={markings, mark=at position .75 with {\arrow{latex}} }, postaction={decorate}] ($(db)$) -- ++(-45:1);
    \draw[double,decoration={markings, mark=at position .75 with {\arrow{latex}} }, postaction={decorate}] (-3,0) -- ++(1,0);

  \end{tikzpicture}

}

\newcommand{\qqcuttwoloops}{

  \begin{tikzpicture} [scale=0.5]
    \path[ use as bounding box] (-2,-4) rectangle (7,4);
    \tikzstyle{ellstyle}=[ fill=white,rounded corners=4pt,inner sep = 1pt, outer sep=1pt];
    \coordinate (fcentre) at (-1,0);
    \coordinate (acentre) at ($({7*cos(30)-sqrt(2)},0)$);
    \coordinate (aa) at ($(fcentre)+(45:1)$);
    \coordinate (ab) at ($(fcentre)+(-45:1)$);
    \coordinate (ba) at ($(acentre)+(135:1)$);
    \coordinate (bb) at ($(acentre)+(225:1)$);
    \coordinate (da) at ($(acentre)+(45:1)$);
    \coordinate (db) at ($(acentre)+(-45:1)$);

    \draw[name path= propright,decoration={markings, mark=at position
      .75 with {\arrow{latex}}}, postaction={decorate}] (aa) to [out =
    45, in = 135, looseness = 1] (ba);

    \draw[name path= propleft,decoration={markings, mark=at position
      .75 with {\arrow{latex}}}, postaction={decorate}] (ab) to [out =
    -45, in = 225, looseness = 1] (bb);

    \draw[fill=gray!40] (fcentre) circle (1);
    \draw[fill=gray!40] (acentre) circle (1);
    \draw[fill=white] (acentre) circle (.6);

    \draw[name path = dashedcut,dashed] ($(aa)!0.5!(ba)+(0,1.5)$) -- ($(ab)!0.5!(bb)+(0,-1.5)$);

    \path [name intersections={of=propleft and dashedcut}] ;
    \coordinate (intleft) at (intersection-1);

    \path [name intersections={of=propright and dashedcut}] ;
    \coordinate (intright) at (intersection-1);

    \node(F) at (fcentre) {$F$};
    \node[above] (qq) at (-2.5,0) {$q$};

    \node[above right] (phA) at ($(da)+(45:1)$) {$\phi^{4}(p_2)$};
    \node[below right] (barph4) at ($(db)+(315:1)$) {$\bar\phi_{A}(p_1)$};

    \node[above right, ellstyle] (ellbir) at (intright) {$\ell_1$};
    \node[below right, ellstyle] (elliki) at (intleft) {$\ell_2$};

    \draw[decoration={markings, mark=at position .75 with {\arrow{latex}} }, postaction={decorate}] ($(da)$) -- ++(45:1);
    \draw[decoration={markings, mark=at position .75 with {\arrow{latex}} }, postaction={decorate}] ($(db)$) -- ++(-45:1);
    \draw[double,decoration={markings, mark=at position .75 with {\arrow{latex}} }, postaction={decorate}] (-3,0) -- ++(1,0);

  \end{tikzpicture}

}

\newcommand{\qqtriplecut}{

  \begin{tikzpicture} [scale=0.5]
    \path[ use as bounding box] (-2,-4) rectangle (7,4);
    \tikzstyle{ellstyle}=[ fill=white,rounded corners=4pt,inner sep = 1pt, outer sep=1pt];
    \coordinate (fcentre) at (-1,0);
    \coordinate (acentre) at ($({7*cos(30)-sqrt(2)},0)$);
    \coordinate (aa) at ($(fcentre)+(45:1)$);
    \coordinate (ab) at ($(fcentre)+(-45:1)$);
    \coordinate (ba) at ($(acentre)+(135:1)$);
    \coordinate (bb) at ($(acentre)+(225:1)$);
    \coordinate (da) at ($(acentre)+(45:1)$);
    \coordinate (db) at ($(acentre)+(-45:1)$);

    \draw[name path= propright,decoration={markings, mark=at position .75 with {\arrow{latex}}}, postaction={decorate}] (aa) to [out = 45, in = 135, looseness = 1] (ba);

    \draw[name path= propleft,decoration={markings, mark=at position .75 with {\arrow{latex}}}, postaction={decorate}] (ab) to [out = -45, in = 225, looseness = 1] (bb);

    \draw[name path= propmid,decoration={markings, mark=at position .75 with {\arrow{latex}}}, postaction={decorate}] ($(fcentre)+(1,0)$) to  ($(acentre)+(-1,0)$);

    \draw[fill=gray!40] (fcentre) circle (1);
    \draw[fill=gray!40] (acentre) circle (1);

    \draw[name path = dashedcut,dashed] ($(aa)!0.5!(ba)+(0,1.5)$) -- ($(ab)!0.5!(bb)+(0,-1.5)$);

    \path [name intersections={of=propleft and dashedcut}] ;
    \coordinate (intleft) at (intersection-1);

    \path [name intersections={of=propmid and dashedcut}] ;
    \coordinate (intmid) at (intersection-1);

    \path [name intersections={of=propright and dashedcut}] ;
    \coordinate (intright) at (intersection-1);

    \node(F) at (fcentre) {$F$};
    \node[above] (qq) at (-2.5,0) {$q$};

    \node[above right] (phA) at ($(da)+(45:1)$) {$\phi^{A}(p_2)$};
    \node[below right] (barph4) at ($(db)+(315:1)$) {$\bar\phi_{4}(p_1)$};

    \draw[decoration={markings, mark=at position .75 with {\arrow{latex}} }, postaction={decorate}] ($(da)$) -- ++(45:1);
    \draw[decoration={markings, mark=at position .75 with {\arrow{latex}} }, postaction={decorate}] ($(db)$) -- ++(-45:1);
    \draw[double,decoration={markings, mark=at position .75 with {\arrow{latex}} }, postaction={decorate}] (-3,0) -- ++(1,0);

    \node(eqzero) at ($(acentre)+(5,0)$) {$=0$};

  \end{tikzpicture}

}

\newcommand{\nonplanar}{
  \begin{tikzpicture}[scale=0.5,cross/.style={path picture={
  \draw[black]
(path picture bounding box.south east) -- (path picture bounding box.north west) (path picture bounding box.south west) -- (path picture bounding box.north east);
}}]

\tikzstyle{qstyle}=[ fill=white,rounded corners=4pt,inner sep = 1pt, outer sep=8pt];

    \draw[decoration={markings, mark=at position 0.5 with {\arrow{latex}}
    }, postaction={decorate}] (0,-3) -- node [midway, below right, inner sep=1pt]{}  (3,0);
    \draw[decoration={markings, mark=at position 0.5 with {\arrow{latex}}
    }, postaction={decorate}] (0,3) -- node [midway, above right, inner sep=1pt]{$\ell_4$} (3,0);

    \draw[decoration={markings, mark=at position 0.5 with {\arrow{latex}}
    }, postaction={decorate}] (0,-3) -- node [midway,below left, inner sep=1pt]{$\ell_3$} (-3,0);
    \draw[decoration={markings, mark=at position 0.5 with {\arrow{latex}}
    }, postaction={decorate}] (0, 3) --  node [midway,above left, inner sep=1pt]{$\ell_6$} (-3,0);
    \draw[decoration={markings, mark=at position 0.5 with {\arrow{latex}} }, postaction={decorate}] (-3,0) -- (-4,0);
    \draw[decoration={markings, mark=at position 0.5 with {\arrow{latex}} }, postaction={decorate}] (3,0) -- (4,0);
    \draw[decoration={markings, mark=at position 0.5 with {\arrow{latex}}
    }, postaction={decorate}] (0,0) -- node [midway,right,inner sep=2pt]{$\ell_1$} (0,-3);
    \draw[decoration={markings, mark=at position 0.5 with {\arrow{latex}}
    }, postaction={decorate}] (0,0) -- node [midway,right,inner sep=2pt]{$\ell_2$} (0,3);

    \draw[decoration={markings, mark=at position 0.75 with {\arrow{latex}} }, postaction={decorate},double]  (-1,0) node [below] {$q$} -- (0,0);

    \node [above] (p1) at (-4,0) {$p_1$};
    \node [above] (P2) at (4,0) {$p_2$};

  \end{tikzpicture}
  }

\newcommand{\nonplanarAcut}{
  \begin{tikzpicture}[scale=0.5,cross/.style={path picture={
  \draw[black]
(path picture bounding box.south east) -- (path picture bounding box.north west) (path picture bounding box.south west) -- (path picture bounding box.north east);
}}]

\tikzstyle{qstyle}=[ fill=white,rounded corners=4pt,inner sep = 1pt, outer sep=8pt];

    \draw[decoration={markings, mark=at position 0.5 with {\arrow{latex}}
    }, postaction={decorate}] (0,-3) --  node [midway,below left, inner sep=1pt]{$\ell_3$} (-3,0);
    \draw[decoration={markings, mark=at position 0.5 with {\arrow{latex}} }, postaction={decorate}] (0, 3) -- (-3,0);

    \draw[decoration={markings, mark=at position 0.5 with {\arrow{latex}}
    }, postaction={decorate}] (0,-3) -- node [midway,below right, inner sep=1pt]{$\ell_5$} (3,0);
    \draw[decoration={markings, mark=at position 0.5 with {\arrow{latex}} }, postaction={decorate}] (0, 3) -- (3,0);

    \draw[decoration={markings, mark=at position 0.5 with {\arrow{latex}} }, postaction={decorate}] (-3,0) -- (-4,0);
    \draw[decoration={markings, mark=at position 0.5 with {\arrow{latex}} }, postaction={decorate}] (3,0) -- (4,0);
    \draw[decoration={markings, mark=at position 0.5 with {\arrow{latex}} }, postaction={decorate}] (0,0) -- (0,-3);
    \draw[decoration={markings, mark=at position 0.5 with {\arrow{latex}}
    }, postaction={decorate}] (0,0) -- node [midway,right,inner sep=2pt]{$\ell_2$} (0,3);

    \draw[dashed, rounded corners=1.5cm] (-3,-1) -- (0,2) -- (3,-1);

    \draw[decoration={markings, mark=at position 0.75 with {\arrow{latex}} }, postaction={decorate},double] (-1,0) node [below] {$q$} -- (0,0);

    \node [above] (p1) at (-4,0) {$p_1$};
    \node [above] (P2) at (4,0) {$p_2$};

  \end{tikzpicture}
}

\newcommand{\nonplanarBcut}{
  \begin{tikzpicture}[scale=0.5,cross/.style={path picture={
  \draw[black]
(path picture bounding box.south east) -- (path picture bounding box.north west) (path picture bounding box.south west) -- (path picture bounding box.north east);
}}]

\tikzstyle{qstyle}=[ fill=white,rounded corners=4pt,inner sep = 1pt, outer sep=8pt];

    \draw[decoration={markings, mark=at position 0.5 with {\arrow{latex}}
    }, postaction={decorate}] (0,-3) -- (3,0);
    \draw[decoration={markings, mark=at position 0.5 with {\arrow{latex}}
    }, postaction={decorate}] (0,3) -- node [midway, above right, inner sep=1pt]{$\ell_5$} (3,0);

    \draw[decoration={markings, mark=at position 0.5 with {\arrow{latex}}
    }, postaction={decorate}] (0,-3) -- (-3,0);

    \draw[decoration={markings, mark=at position 0.5 with {\arrow{latex}}
    }, postaction={decorate}] (0, 3) -- node [midway,above left, inner sep=1pt]{$\ell_3$} (-3,0);

    \draw[decoration={markings, mark=at position 0.5 with {\arrow{latex}} }, postaction={decorate}] (-3,0) -- (-4,0);
    \draw[decoration={markings, mark=at position 0.5 with {\arrow{latex}} }, postaction={decorate}] (3,0) -- (4,0);
    \draw[decoration={markings, mark=at position 0.5 with {\arrow{latex}}
    }, postaction={decorate}] (0,0) -- node [midway,right,inner sep=2pt]{$\ell_2$} (0,-3);
    \draw[decoration={markings, mark=at position 0.5 with {\arrow{latex}}
    }, postaction={decorate}] (0,0) -- (0,3);

    \draw[dashed, rounded corners=1.5cm] (-3,1) -- (0,-2) -- (3,1);

    \draw[decoration={markings, mark=at position 0.75 with {\arrow{latex}} }, postaction={decorate},double] (-1,0) node [above] {$q$} -- (0,0);

    \node [above] (p1) at (-4,0) {$p_1$};
    \node [above] (P2) at (4,0) {$p_2$};

  \end{tikzpicture}
  }

\newcommand{\laddertriangle}{
  \begin{tikzpicture} [scale=0.5]
    \tikzstyle{ellstyle}=[ fill=white,rounded corners=4pt,inner sep = 1pt, outer sep=4pt];
    \path[ use as bounding box] (-4,2) rectangle (4,-7);
    \draw[double,decoration={markings, mark=at position .75 with {\arrow{latex}} }, postaction={decorate}] (0,1) -- node[left] {$q$} ++(-90:1) ;
    \draw[decoration={markings, mark=at position .5 with {\arrow{latex}} }, postaction={decorate}] (0,0) -- node [midway, right, inner sep=1pt, outer sep=4pt]{$\ell_1$} (300:3);
    \draw (0,0) -- (240:3);
    \draw[decoration={markings, mark=at position .5 with {\arrow{latex}} }, postaction={decorate}] (240:3) -- node [midway, left, inner sep=1pt, outer sep=4pt]{$\ell_4$} (240:6);
    \draw[decoration={markings, mark=at position .5 with {\arrow{latex}} }, postaction={decorate}] (300:3) -- node [midway, right, inner sep=1pt, outer sep=4pt]{$\ell_3$} (300:6);
    \draw[decoration={markings, mark=at position .5 with {\arrow{latex}} }, postaction={decorate}] (240:6) -- (240:7);
    \draw[decoration={markings, mark=at position .5 with {\arrow{latex}} }, postaction={decorate}] (300:6) -- (300:7);
    \draw[decoration={markings, mark=at position .5 with {\arrow{latex}} }, postaction={decorate}] (300:3) -- node [midway, below, inner sep=1pt, outer sep=4pt]{$\ell_5$} (240:3);
    \draw[decoration={markings, mark=at position .5 with {\arrow{latex}} }, postaction={decorate}] (300:6) -- node [midway, below, inner sep=1pt, outer sep=4pt]{$\ell_6$} (240:6);
    \begin{scope}[node distance=5]
v
    \node[below left] (bir) at (240:7) {$p_1$};
    \node[below right] (iki) at (300:7) {$p_2$};
    \end{scope}
  \end{tikzpicture}
}

\newcommand{\crooked}{
  \begin{tikzpicture} [scale=0.5]
    \tikzstyle{ellstyle}=[ fill=white,rounded corners=4pt,inner sep = 1pt, outer sep=4pt];
    \path[ use as bounding box] (-4,2) rectangle (4,-7);
    \draw[double,decoration={markings, mark=at position .75 with {\arrow{latex}} }, postaction={decorate}] (0,1) -- node[left] {$q$} ++(-90:1) ;
    \draw[decoration={markings, mark=at position .5 with {\arrow{latex}} }, postaction={decorate}] (0,0) -- node [midway, right, inner sep=1pt, outer sep=4pt]{$\ell_1$} (300:3);
    \draw (0,0) -- (240:6);
    \draw[decoration={markings, mark=at position .5 with {\arrow{latex}} }, postaction={decorate}] (300:3) -- node [midway, right, inner sep=1pt, outer sep=4pt]{$\ell_3$} (300:6);
    \draw[decoration={markings, mark=at position .5 with {\arrow{latex}} }, postaction={decorate}] (240:6) -- (240:7);
    \draw[decoration={markings, mark=at position .5 with {\arrow{latex}} }, postaction={decorate}] (300:6) -- (300:7);
    \draw[decoration={markings, mark=at position .5 with {\arrow{latex}} }, postaction={decorate}] (300:3) -- node [midway, above, inner sep=1pt, outer sep=4pt]{$\ell_5$} (240:6);
    \draw (240:6) -- (300:6);
    \begin{scope}[node distance=5]

    \node[below left] (bir) at (240:7) {$p_1$};
    \node[below right] (iki) at (300:7) {$p_2$};

    \end{scope}
  \end{tikzpicture}
}

\newcommand{\fan}{
  \begin{tikzpicture} [scale=0.5]
    \tikzstyle{ellstyle}=[ fill=white,rounded corners=4pt,inner sep = 1pt, outer sep=4pt];
    \path[use as bounding box] (-4,2) rectangle (4,-7);
    \draw[double,decoration={markings, mark=at position .75 with {\arrow{latex}} }, postaction={decorate}] (0,1) -- node[left] {$q$} ++(-90:1) ;
    \draw[decoration={markings, mark=at position .5 with
      {\arrow{latex}} }, postaction={decorate}] (0,0) -- node[left]
    {$\ell_1$} (240:6);
    \draw[decoration={markings, mark=at position .5 with
      {\arrow{latex}} }, postaction={decorate}] (0,0) -- node[right]
    {$\ell_3$} (300:6);
    \draw (0,0) -- ($ (240:6) !.5! (300:6) $);
    \draw (240:6)[decoration={markings, mark=at position .5 with
      {\arrow{latex}} }, postaction={decorate}] -- node[below] {$\ell_5$} ($ (240:6) !.5! (300:6) $);
    \draw (300:6) -- ($ (240:6) !.5! (300:6) $);
    \draw[decoration={markings, mark=at position .5 with
      {\arrow{latex}} }, postaction={decorate}] (240:6) --  (240:7) node[below left] {$p_1$};
    \draw[decoration={markings, mark=at position .5 with
      {\arrow{latex}} }, postaction={decorate}] (300:6) --  (300:7) node[below right] {$p_2$};
  \end{tikzpicture}
}

\newcommand{\sunset}{
  \begin{tikzpicture} [scale=0.5]
    \draw (0,0) circle (1) ;
    \draw (-1.5,0) -- (1.5,0);
  \end{tikzpicture}
}
\newcommand{\tri}{
  \begin{tikzpicture} [scale=0.5]
    \draw (0,0) circle (1) ;
    \draw (0,-1) -- (0,1);
    \draw (-1.5,0) -- (-1,0);
    \draw (0,1) -- (1,1);
    \draw (0,-1) -- (1,-1);
  \end{tikzpicture}
}
\newcommand{\glass}{
  \begin{tikzpicture} [scale=0.5]
    \draw (0,0) circle (1) ;
    \draw (2,0) circle (1) ;
    \draw (-1.5,0) -- (-1,0);
    \draw (3,0) -- (3.5,0);

  \end{tikzpicture}
}

\newcommand{\trianx}{
  \begin{tikzpicture} [scale=0.5]
    \draw (20:1) -- (-20:3);
    \draw[white, line width=4pt] (-20:1) -- (20:3);
    \draw (-0.5,0) -- (0,0);
    \draw (0,0) -- (20:3.5);
    \draw (0,0) -- (-20:3.5);
    \draw (-20:1) -- (20:3);
  \end{tikzpicture}
}

\newcommand{\fourpdoublelinetypea}[4]{
  \begin{tikzpicture}
    \draw[name path= propright,decoration={markings, mark=at position .5 with {\arrow{>}}}, postaction={decorate}] (-0.95,1.05) node
    [above left] {$#4$} to [out = -45, in = -135, looseness = 1] (0.95,1.05) node [above right] {$#1$};
    \draw[name path= propright,decoration={markings, mark=at position .5 with {\arrow{>}}}, postaction={decorate}] (0.95,-1.05) node [below right] {$#2$} to [out = 135, in = 45, looseness = 1] (-0.95,-1.05) node [below left] {$#3$};
    \draw[name path= propright,decoration={markings, mark=at position .5 with {\arrow{>}}}, postaction={decorate}, dashed] (1.05,0.95)  to [out = -135, in = 135, looseness = 1] (1.05,-0.95) ;
    \draw[name path= propright,decoration={markings, mark=at position .5 with {\arrow{>}}}, postaction={decorate}, dashed] (-1.05,-0.95)  to [out = 45, in = -45, looseness = 1] (-1.05,0.95) ;

  \end{tikzpicture}
}

\newcommand{\fourpdoublelinetypeb}[4]{
  \begin{tikzpicture}
    \draw[name path= propright,decoration={markings, mark=at position .5 with {\arrow{<}}}, postaction={decorate},dashed]  (-0.95,1.05) node
    [above left] {$#4$} to [out = -45, in = -135, looseness = 1] (0.95,1.05) node [above right] {$#1$}  ;
    \draw[name path= propright,decoration={markings, mark=at position .5 with {\arrow{<}}}, postaction={decorate},dashed] (0.95,-1.05) node [below right] {$#2$} to [out = 135, in = 45, looseness = 1] (-0.95,-1.05) node [below left] {$#3$};
    \draw[name path= propright,decoration={markings, mark=at position .5 with {\arrow{<}}}, postaction={decorate}] (1.05,0.95)  to [out = -135, in = 135, looseness = 1] (1.05,-0.95) ;
    \draw[name path= propright,decoration={markings, mark=at position .5 with {\arrow{<}}}, postaction={decorate}] (-1.05,-0.95)  to [out = 45, in = -45, looseness = 1] (-1.05,0.95) ;

  \end{tikzpicture}
}

\newcommand{\ang}[1]
{
  \langle #1 \rangle
}

\newcommand{\sqr}[1]
{
  [ #1 ]
}
\newcommand{\tlambda}
{
  \tilde{\lambda}
}

\newcommand{\diffd}{\mathrm{d}}

\newcommand{\mmbox}[3]{\makebox[#1][#2]{#3}}

\newcommand*\circled[1]{\tikz[baseline=(char.base)]{
            \node[shape=circle,draw,inner sep=0.5pt] (char) {\tiny #1};}}

\font\mybb=msbm10 at 12pt
\def\bb#1{\hbox{\mybb#1}}

\font\myBB=msbm10 at 18pt
\def\BB#1{\hbox{\myBB#1}}

\newcommand\note[1]{\mbox{}\marginpar{\footnotesize\raggedright\hspace{0pt}\color{blue}\emph{#1}}}

\begin{document}
\begin{flushright}
%ArXiv:yymm.nnnn \\
QMUL-PH-12-23
\end{flushright}

\vspace{20pt}

\begin{center}

  {\Large \bf    Two-loop Sudakov Form Factor in ABJM}
\\
\vspace{11pt}
\vspace{32pt}
{\mbox {\bf A.~Brandhuber, \"O.~G\"urdo\u{g}an, D.~Korres, R.~Mooney and G.~Travaglini}}%
\footnote{ {\sffamily \{\tt a.brandhuber, o.c.gurdogan, d.korres,
r.j.b.mooney,  g.travaglini\}@qmul.ac.uk }}

\bigskip

{\em Centre for Research in String Theory\\
School of Physics and Astronomy\\
Queen Mary University of London\\
Mile End Road, London, E1 4NS, UK
 }

\bigskip

\vspace{30pt} {\bf Abstract}

\end{center}

\noindent 
We compute the two-loop Sudakov form factor in three-dimensional
$\mathcal{N} = 6$ superconformal Chern-Simons theory, using generalised unitarity. 
As an intermediate step, we derive the non-planar part of the one-loop four-point amplitude in terms of box integrals. 
Our result for the Sudakov form factor is given by a single non-planar tensor integral with  uniform degree of transcendentality, 
and  is in agreement with the known infrared divergences of two-loop amplitudes in ABJM theory. 
We also discuss a number of interesting properties satisfied by related  three-dimensional integral functions. 

\setcounter{page}{0}
\thispagestyle{empty}
\newpage

%%%%%%%%%%%%%%%%%%%%%%%%%%%%%%%%%%%%%%%%%%%%%%%%%%%%%%%%%%%%%%%%%

\tableofcontents

\setcounter{footnote}{0}

 \section{Introduction}

In this paper we continue our investigation of amplitudes and form factors
in three-dimensional $\cN=6$ Chern-Simons matter theory, also known as ABJM  \cite{abjm}. 
This theory is closely related to the maximally supersymmetric theories  constructed in
\cite{bl,gustavsson}, and provides an interesting example of holographic duality in three dimensions.

One interesting fact about ABJM theory is the presence of very similar, hidden
structures  as in $\cN=4$ super Yang-Mills (SYM). In particular, dual conformal symmetry
\cite{Huang:2010qy}, integrability, \cite{Minahan:2008hf,Bak:2008cp, Lipstein:2011ej}, duality with Wilson loops at four points
\cite{Henn:2010ps,Wiegandt:2011uu,silviacorto},
uniform transcendentality of the two-loop four-point  \cite{Chen:2011vv,
Bianchi:2011dg} and six-point amplitude \cite{Huang:2012hr},
colour-kinematics duality \cite{Bargheer:2012gv} have made their appearance in both theories. Furthermore, several
powerful methods to compute four-dimensional amplitudes could be adapted and generalised  to three dimensions -- BCFW recursion relations \cite{Gang:2010gy}, generalised unitarity \cite{Huang:2010qy, Bargheer:2012cp, btw1, btw2}, 
and Grassmannian formulations \cite{lee} of amplitudes.

With the broad aim of further exploring these similarities between ABJM and $\cN=4$ SYM, we initiate here a study of form factors in ABJM. In the present paper we focus on Sudakov form factors of half-BPS operators, i.e.~the overlap of a state created by an operator built from two scalars and a two-particle on-shell state.
In $\cN=4$ SYM,   Sudakov form factors were first studied (at one and two loops) by van Neerven  in \cite{vanneerven}. More recently, these quantities were revisited and various generalisations and extensions were achieved: form factors with more than two external on-shell particles were considered in \cite{bsty, bork1, harmony, bork2} at one loop, and BPS operators with more than two scalars were studied in \cite{bork1}. In \cite{bty} the two-loop, three-point from factor was calculated using generalised unitarity and, alternatively, from symbology \cite{gsvv}, and a remarkable link to Higgs plus one-jet amplitudes in QCD \cite{gehrmann} was unearthed, providing a new example of the principle of maximal transcendentality, first observed in  \cite{Kotikov:2004er} for anomalous dimensions of composite operators.
The calculation of the Sudakov form factor was pushed to three loops in \cite{ghh}, where it was also found that the result could be expressed in terms of a set of integral functions that individually are uniformly transcendental, and more recently its  four-loop integrand was constructed in \cite{boels} using colour-kinematics duality \cite{bcj}. In parallel there have been studies of form factors at strong coupling using AdS/CFT duality and integrability, initially in \cite{mz} and more recently in \cite{gaoyang} where  insertions of multiple operators were also considered.

As alluded to above, these studies revealed that much of the technology and mathematical structures known from amplitudes also apply to form factors, e.g.~recursion relations and unitarity, integrability and Y-systems, maximal transcendentality, symbology, colour-kinematics duality. However, there are also marked differences, in particular the appearance of non-planar integral topologies and the absence of dual conformal/Yangian symmetry.

In this paper we find that the result for the two-loop Sudakov form factor in ABJM has uniform degree of transcendentality and captures correctly the infrared divergences of two-loop amplitudes. The slightly unusual observation is that it is expressed in terms of single non-planar integral function with a very peculiar numerator. This is different from the story in $\cN=4$ SYM, where the two-loop form factor \cite{vanneerven} is expressed in terms of a planar and a non-planar integral function which both have trivial numerators and are separately transcendental.
The role of this special numerator is two-fold: it removes unphysical infrared divergences linked to internal three-particle vertices, and at the same time makes the result transcendental. We have investigated these issues for several planar two-loop topologies with very similar findings, i.e.~choosing the numerators in such a way to remove unphysical infrared divergences makes these integral transcendental. These planar topologies do not appear in the ABJM form factor but are ingredients of the form factors in ABJ which is also transcendental.

There are some parallels with $\cN=4$ SYM, namely there exist certain superficially dual-conformal integrands which upon integration lead to divergent results even with external massive kinematics \cite{Drummond:2007aua}. There, this phenomenon was observed only at higher loops and it was shown that such integrals, which could potentially contribute to amplitudes, actually have  vanishing coefficient. Similar observations were made in the study of two-loop amplitudes in ABJM \cite{Huang:2012hr}, where important constraints on integral functions were inferred from the condition of vanishing triple cuts involving three-particle (and five-particle) amplitudes. 
It is the vanishing triple cuts, involving three-particle amplitudes, that we identify to guarantee the absence of unphysical infrared divergences  and of terms which would spoil the uniform transcendentality of the result. Furthermore, these triple cuts  give powerful constraints and cross-checks on the (numerators of) integral functions. This observation also points to more general questions in the study of amplitudes and form factors in three and four dimensions: what is a good basis of integral functions? What are the physical and practical criteria for the choice? What are the properties of this basis? Some of the desirable features would be that the expansion coefficients are $\epsilon$-independent (in the case of ABJM this requires tensor integrals instead of scalar integrals) and that the integral functions individually are transcendental%
\footnote{See \cite{henn} for a recent proposal in four dimensions.}, but it would clearly be interesting to understand better the underlying physical and mathematical criteria for any potential choice. The ABJ(M) form factors seem to provide an interesting playground to study such issues.

The rest of the paper is organised as follows. After  reviewing some general properties of ABJM amplitudes in Section 2, we present in Section 3 the calculation of the non-planar part of the one-loop four-point amplitude in this theory. The complete --  planar plus non-planar --  integrand of the four-point amplitude is a key ingredient in the construction of the Sudakov form factor, which is discussed in Section 4. There, we  derive this  quantity firstly at one and then at two loops, using two- and three-particle cuts at the level of the integrand. As mentioned earlier, the result is expressed in terms of a single non-planar integral topology with a special numerator, whose properties we discuss in detail. In particular, we consider certain three-vertex cuts, which put strong constraints on the form of these numerators. We also compare our result to the known infrared divergences of ABJM amplitude at two loops, finding complete agreement. Finally, in Section 5 we introduce three planar integral topologies which contribute to the ABJ form factor. We discuss their properties and present their maximally transcendental result. Three appendices contain details on half-BPS operators in ABJM,  on the one-loop box function in terms of which the four-point amplitude is expressed, and on the reduction to master integrals of the integral topologies discussed in this paper. 

 %%%%%%%%%%%%%%%%%%%%%%%%%%%%%%%%%%%%%%%%%%%%%%%%%%%%%%%%%%%%%%%%%

\section{Scattering amplitudes in ABJM theory}

In the following we briefly review some key facts of ABJM theory, and in
particular of its tree amplitudes, which appear in the construction of
loop amplitudes and form factors using unitarity \cite{bddk,fusing,bdkgen,bcfgen}.%
\footnote{In this paper we
  follow the conventions of Section 2 and Appendix A of \cite{btw1} for the
  ABJM  superamplitudes and the three-dimensional spinor helicity
  formalism,
respectively.}

\subsection{Superamplitudes}

Three-dimensional $\cN=6$ Chern-Simons matter theory \cite{abjm} (or, in short, ABJM)
is a quiver theory with gauge group $U_k(N) \times U_{-k}(N)$, where $k$
and $-k$ are the Chern-Simons levels of the gauge fields $A_\mu$ and
$\hat{A}_\mu$, respectively.  The matter fields comprise four complex
scalars $\phi^A$ and four fermions $\psi^\alpha_A$, where  $A=1, \ldots, 4$
is a $SU(4)$ $R$-symmetry index and $\alpha=1,2$ is a spin index.  The
fields $(\phi^A)^{i}_{\bar j}$ and $(\psi^\alpha_A)^{i}_{\bar j}$ transform
in the bifundamental representation $(N, \bar{N})$ of the  gauge group,
while $(\bar\phi_A)^{\bar i}_j$ and $(\bar\psi_\alpha^A)^{\bar i}_j$
transform in the $(\bar{N}, N)$, with $i, \bar{j} = 1, \ldots , N$. An
interesting variant of ABJM is the so-called ABJ theory, i.e.~$\cN=6$
Chern-Simons theory with  gauge group $U_k(N) \times U_{-k}(N^\prime)$. In
this case $i =  1, \ldots , N$, $\bar{j} =1, \ldots , N^\prime$.  Note that
in the ABJ(M) theory the gauge fields are non-dynamical because of  the
topological nature of the Chern-Simons action, and hence they cannot appear
as external states.

The momenta of the particles can be written efficiently in the
three-dimensional spinor helicity formalism as
\beq
p_{\alpha \beta} \ := \ \lambda_\a \lambda_\b\, ,
 \eeq
where $\lambda_\a$ are commuting spinors.  The states of the ABJM theory
can be packaged into two Nair superfields \cite{Nair, Huang:2010rn},
\begin{align}
  \Phi (\l, \eta) & =
  \phi^4(\l) + \eta^A \psi_A(\l) + \frac{1} {2} \e_{ABC} \eta^A \eta^B
  \phi^C(\l) + \frac{1}{3!} \e_{ABC} \eta^A \eta^B \eta^C \psi_4(\l) \, ,
  \\
  \bar{\Phi} (\l , \eta) & =
  \bar{\psi}^4(\l) + \eta^A \bar{\phi}_A(\l)
  + \frac{1} {2} \e_{ABC} \eta^A \eta^B \bar{\psi}^C(\l) + \frac{1} {3!}
  \e_{ABC} \eta^A \eta^B \eta^C \bar{\phi}_4(\l) \, ,
\end{align}
where $\eta^A$, $A=1,2,3$ are Grassmann coordinates parameterising an
$\mathcal{N}=3$ superspace. The superfields $\Phi$ and $\bar{\Phi}$ carry
colour indices $\Phi^i_{\bar{j}}$ and $\bar{\Phi}^{\bar{i}}_j$. Note that
$\Phi$ is bosonic while $\bar\Phi$ is fermionic.  This description breaks
the $SU(4)$ $R$-symmetry of the theory
down to a manifest $U(3)$.

Colour-ordered partial amplitudes were introduced in
\cite{Bargheer:2010hn}, and we denote them as $\cA( \bar\Phi_1, \Phi_2,
\ldots , \Phi_n)$. An important feature of ABJ(M)  is that  any amplitude
with an odd number of particles vanishes, as a simple consequence of gauge
invariance.  Invariance under translations and supersymmetry
transformations ensures that amplitudes are proportional to $\delta^{(3)}
(P) \delta^{(6)} ( Q)$, where $Q_\alpha^A$ and $P_{\alpha \beta}$ are the
total momentum and supermomentum  of $n$ particles, respectively:
\begin{equation}
  P_{\alpha \beta} \, := \, \sum_{i = 1}^n \lambda_{i, \alpha} \lambda_{i, \beta} \ , \qquad
  Q_\alpha^A
  \, := \,
  \sum_{i = 1}^n\l_{i, \a} \eta^A_i
  \, .
\end{equation}
The first non-vanishing amplitude of the theory occurs at four points, and
is the basic building block of higher-point amplitudes.  At
tree level it is given by the following compact expression
\cite{Agarwal:2008pu},
\begin{equation}
  \label{eq:amptree}
  \mathcal{A}_4^
  \text{(0)}
  \bigl( \bar1, 2, \bar3, 4\bigr)
  \ = \
  i \, \frac{\delta^{(6)} (Q) \delta^{(3)} (P)} {\ang{1 \, 2} \, \ang{2 \, 3}}
  \ .
\end{equation}
As usual, component amplitudes can be obtained by extracting the
coefficient of the appropriate monomial in the $\eta_i$ variables. For
instance, in order to pick the component amplitude
$A \big( \bar\phi_A(p_1),   \phi^4(p_2), \bar\phi_4(p_3), \phi^A(p_4)
\big)$ we need to expand the fermionic delta function $\delta^{(6)} ( Q)$
and keep the term
$(\eta_1)^1 (\eta_2)^0 (\eta_3)^3 (\eta_4)^2 \ang{1 \, 3} \ang{3 \, 4}^2 $.

  \begin{figure}
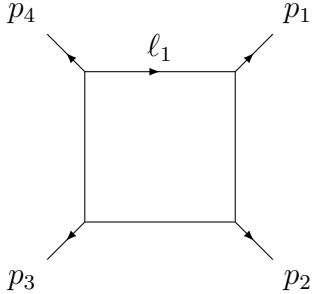

  \centering
  \boxdiagram
 \caption{\it The one-loop box function in  \eqref{YTbox}.}
\end{figure}
  \label{sec:facts-abjm-theory}

The one-loop colour-ordered four-point  superamplitude%
\footnote{ Here, and in what follows, we use the normalisation $1/(i \pi^{D/2})$
  per loop. In \cite{Chen:2011vv}, the normalisation is $1/(2 \pi)^D$.  }%
~was constructed in
\cite{Chen:2011vv}, and is equal to%
\footnote{We suppress the Chern-Simons level $k$, which will be reinstated
at the end of our two-loop calculation.}
\begin{equation}
    \label{eq:amponeloop} \mathcal{A}^\text{(1)}  \bigl( \bar1, 2, \bar3,
    4\bigr) \ =\ i \mathcal{A}^\text{(0)}  \bigl( \bar1, 2, \bar3, 4\bigr)
    \,N\,  I(1,2,3,4) \ ,
    \eeq
where   $s_{ij} = (p_i + p_j)^2$.  The
one-loop integral $I(1,2,3,4)$ is defined by
\beq
\label{YTbox}
I(1,2,3,4) \ := \   \int\!\!\frac{d^D \ell} {i \pi^{D/2}}\, \frac{
        s_{12} \,\Tr (\ell  \, p_1 \, p_4)  \, + \,   \ell^2 \, \Tr (p_1  \, p_2 \,  p_4) }
  {\ell^2 (\ell - p_1)^2(\ell - p_1 - p_2)^2(\ell + p_4)^2} \, ,
\end{equation}
with $D=3 - 2\eps$.
Note that $\Tr (abc)  = 2  \epsilon(a, b, c):=2 \epsilon_{\mu \nu \rho}
a^\mu b^\nu c^\rho$.

Explicit evaluation of the right-hand side of \eqref{eq:amponeloop} shows
that  $\mathcal{A}^\text{(1)}(\bar1, 2 , \bar3, 4) $ is of
$\mathcal{O}(\eps)$, and hence vanishes in three dimensions
\cite{Chen:2011vv}.  This is consistent with the fact that all one-loop
amplitudes in ABJM can be expanded in terms of one-loop triangle functions
\cite{Bargheer:2012cp}, as expected from dual conformal invariance. The
vanishing of the four-point amplitude then follows since one-mass (and two-mass) triangles vanish
in three dimensions.  Very interestingly, the box
function with the particular numerator in \eqref{YTbox} is also
dual conformal invariant, as was demonstrated in \cite{Chen:2011vv}
using a five-dimensional embedding formalism. Furthermore, the expression
for $\mathcal{A}^\text{(1)} (\bar1, 2, \bar3, 4)$ given in
\eqref{eq:amponeloop}  is correct to all orders in the dimensional
regularisation parameter $\eps$. In the following, the integrand of
\eqref{YTbox} will be a crucial ingredient in applying unitarity at
the integrand level.

%%%%%%%%%%%%%%%%%%%%%%%%%%%%%%%%%%%%%%%%%%%%

\subsection{Colour ordering at tree level}

As is well known from experience in $\cN=4$ SYM, starting from two
loops, cuts of form factors receive contributions from non-planar
amplitudes which are as leading as those arising from the planar amplitudes
(see for example \cite{Gehrmann:2011xn,bty}). For our present
purposes, we will need the complete (planar and non-planar) one-loop
amplitude in ABJM at four points. We  recall that
complete tree amplitudes, denoted here by $\tilde\cA$,
are given by \cite{Bargheer:2010hn,Lipstein:2011ej}
\beq
\tilde\cA (\bar{1}, 2, \ldots, n )
\ = \
\sum_{\cP_n}  {\rm sgn} (\sigma) \cA^{(0)}
\big(
    \sigma(\bar1),  \sigma(2), \sigma(\bar3), \ldots , \sigma(n)
\big) \,
\big[
    \sigma(\bar1),  \sigma(2), \sigma(\bar3), \ldots , \sigma(n)
\big]\, ,
\eeq
where $\cP_n:= (S_{n/ 2} \times S_{n/ 2}) / C_{n/ 2}$  are permutations of
$n$ sites that only mix even (bosonic)  and odd (fermionic) particles among
themselves, modulo cyclic permutations by two sites, and the function ${\rm
sgn }(\sigma)$  is equal to $ -1$ if $\sigma$  involves  an odd permutation
of the odd (fermionic)  sites, and $+1$ otherwise.  $\cA^{(0)} (\bar{1}, 2,
\bar{3}  \ldots , n )$ are colour-ordered tree amplitudes, and
we have also defined
\beq
\label{r1}
\big[ \bar{1},  2, \bar{3}, \ldots , n \big]
\ :=\
\delta^{\bar{i}_1}_{\bar{i}_2}\delta^{i_2}_{i_3}\delta^{\bar{i}_3}_{\bar{i}_4} \cdots \delta^{i_n}_{i_1}\, .
\eeq
In  the following we will just write $\big[1, 2, \cdots, n]$ without
specifying if a particle is barred (i.e.~fermionic) or un-barred
(bosonic), with the understanding that the first entry in the bracket always represents a fermionic field.

\begin{figure}
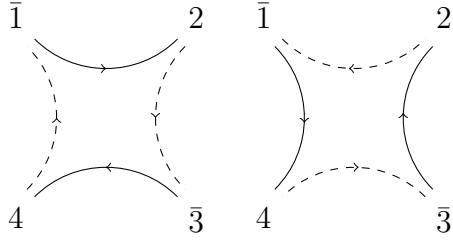

  		\centering
		\fourpdoublelinetypea{2}{\bar{3}}{4} {\bar{1}}
 		\fourpdoublelinetypeb{2}{\bar{3}}{4}{\bar{1}}
		        \caption{\it The two possible colour orderings $[1,2,3,4]$ and $[1,4,3,2]$  appearing in the four-point tree-level amplitude \eqref{c4ptt}.}
        \label{colourstructuresattree}
\end{figure}

As an example, we consider the complete four-point amplitude at tree
level. It includes the two colour structures $[1,2,3,4]$ and $[1,4,3,2]$
(see Figure \ref{colourstructuresattree}) and is given by the following
expression: \beq
\label{c4ptt}
\tilde\cA^{(0)} (\bar{1}, 2, \bar{3}, 4 ) \ = \ \cA^{(0)} (\bar{1}, 2, \bar{3}, 4 )
\Big(
\big[1,2,3,4 \big] \, - \, \big[1,4,3,2] \Big)
\ .
\eeq
We have also used that
\beq
\label{1lsy}
\cA^{(0)} (\bar{1}, 2, \bar{3}, 4 ) \  =\  \cA^{(0)} (\bar{3}, 2, \bar{1}, 4 ) \, ,
\eeq
a fact that follows  from
\eqref{eq:amptree}.

\subsection{Symmetry properties of amplitudes}

It  is useful to recall the following general relations for
colour-ordered amplitudes \cite{Bargheer:2012cp}:
 \beq
\cA^{(l)} (\bar{1}, 2, \bar{3} \cdots, n)
\ = \
(-)^{{n\over2} - 1}\,  \cA^{(l)} (\bar{3}, 4,  \cdots, \bar{1}, 2)\ ,
\eeq
and
\beq
\cA^{(l)} (\bar{1}, 2, \bar{3} \cdots, n)
\ = \
(-)^{{n(n-2)\over8} + l} \, \cA^{(l)} (\bar{1}, n, \overline{n-1}, n-2,  \cdots, \bar{3}, 2)\ .
\eeq
We note that complete amplitudes should behave under exchange of any two
particles as the spin-statistics theorem requires. In particular we expect
\beq
\label{minus}
\tilde\cA^{(l)}  (\bar{1}, 2, \bar{3}, 4)\, = \, -\, \tilde\cA^{(l)} (\bar{3}, 2, \bar{1}, 4 )
\ ,
\eeq
at any loop order. This requires this explains the presence of the crucial minus sign
between the two possible colour structures in \eqref{c4ptt}.

\section{The complete one-loop amplitude}
\subsection{Results}

In this section we present our result for the complete four-point amplitude
at one loop in ABJM.  As mentioned earlier, this amplitude will be needed
in order to construct the two-particle cuts of the two-loop form factor.
The one-loop four-point amplitude  is given by the sum of a planar and
non-planar contribution:
\beq
\label{res1l}
  \tilde\cA^{(1)} (\bar{1}, 2, \bar{3}, 4)
      \ = \
  \cA^{(1)}_{\rm P}  (\bar{1}, 2, \bar{3}, 4)\, + \, \cA^{(1)}_{\rm NP} (\bar{1}, 2, \bar{3}, 4)
\ ,
\eeq
where
\begin{align}
\label{xxxxx}
   \cA^{(1)}_{\rm P} (\bar{1}, 2, \bar{3}, 4 )
    = &
   \, i \, N\, \cA^{(0) }  (\bar{1}, 2, \bar{3}, 4 ) \, I(1,2,3,4)
\Big( \big[1,2,3,4] +  \big[1,4,3,2] \Big) \,   ,
\intertext{and}
\label{np1l}
\begin{split}
\cA^{(1)}_{\rm NP} (\bar{1}, 2, \bar{3}, 4 )
 = &
-2\, i\, \cA^{(0)}  (\bar{1}, 2, \bar{3}, 4)
\Big[  \Big( I(1,2,3,4) -  I(4,2,3, 1)\Big) [1,2][3,4]  \\
& \hspace{3.5cm} - \Big( I(2,3,4,1) -   I(1,3,4,2)\Big) [1,4][3,2] \Big]
\ .
\end{split}
\end{align}
Note that the double-trace structure $[1,2]$ is
\beq
\label{nota}
[1,2] \ = \ \delta^{\bar{i}_1}_{\bar{i}_2} \delta^{i_2}_{i_1}
\ .
\eeq
The complete one-loop amplitude can also be written in the
following way,
\begin{multline}
\negthickspace \negthickspace
\frac{
\tilde{\mathcal{A}}^{(1)}(\bar{1},2,\bar{3},4)}
{\mathcal{A}^{{(0)}}(\bar{1},2,\bar{3},4)}
\!=\!
i \Big\{
  I(1,2,3,4)
  \Big[
    N
   	 \big(
  	  	[1,2,3,4]+[1,4,3,2]
	    \big)
	    - 2[1,2][3,4]-2[1,4][3,2]
  \Big]\\
  + 2 \Big[
     I(4,2,3,1)[1,2][3,4]-I(1,3,4,2)[1,4][3,2]
    \Big]
  \Big\}.
\end{multline}

\subsection{Symmetry properties of the one-loop amplitude}

Before discussing the derivation of \eqref{res1l},  it is  instructive to
prove that $\cA^{(1)}_{\rm P}$ and  $\cA^{(1)}_{\rm NP}$ are antisymmetric
under the swap $\bar{1} \leftrightarrow \bar{3}$ (see \eqref{minus}). In
order to show this one needs to use \eqref{1lsy} and the following
relations satisfied by the one-loop box
\eqref{YTbox}:
\beq
\label{prima}
I(a,b,c,d) \ = \ -\, I(b,c,d,a) \ , \qquad I(a,b,c,d) =
- I(c, b, a, d) \ .
\eeq
These relations  state  that by cyclically shifting the labels of the
external legs of the box function \eqref{YTbox} by one unit one picks a
minus sign; and similarly if one swaps two non-adjacent legs. Both
relations  are straightforward to prove using the definition \eqref{YTbox}
of the box function. One then finds,
 \begin{align}
      \label{ssss}
 I(3,2,1,4)  - I(4,2,1,3)& = I(2,3,4,1) - I(1,3,4,2)
 \ ,
 \nonumber \\
 I(2,1,4,3) - I(3,1,4,2) &= I(1,2,3,4) - I(4,2,3,1)
 \ . \\
 \intertext{Using \eqref{ssss} we get}
 \cA^{(1)}_{\rm P} (\bar{1}, 2, \bar{3}, 4 )
 & =
 - \cA^{(1)}_{\rm P} (\bar{3}, 2, \bar{1}, 4 )  \ ,
 \nonumber \\
 \cA^{(1)}_{\rm NP} (\bar{1}, 2, \bar{3}, 4 )
 &=
 - \cA^{(1)}_{\rm NP} (\bar{3}, 2, \bar{1}, 4 )  \ .
 \end{align}
Notice the presence of a  minus sign between the two non-planar colour
structure $[1,2][3,4]$ and $[1,4][3,2]$ appearing in the non-planar
one-loop amplitude
\eqref{np1l}.

\subsection{Derivation of the complete one-loop amplitude from cuts}

We now briefly outline the strategy for the derivation of the complete
one-loop amplitude \eqref{res1l}, which is very similar to that in $\cN=4$
SYM, see for example \cite{Feng:2011fja}.  We consider the two-particle
cuts of the complete one-loop amplitude, which are obtained by  merging two
tree-level amplitudes summed over all possible colour structures and
internal particle species. We will see that each cut can be re-expressed in
terms of cuts of sums of  box functions \eqref{YTbox}. The sum over
internal species is (partially) performed via an  integration over the
Grassmann variables  $\eta_{\ell_1}$ and $\eta_{\ell_2}$ associated to the
cut momenta. If one of the particles crossing is bosonic and the other is
fermionic we also have to add to this the same expression with $\ell_1
\leftrightarrow \ell_2$ -- this is necessary only for the $s$- and
$t$-cuts.  For instance, the $s$-cut integrand of the one-loop amplitude
is%
\footnote{For convenience we include here a factor of $\frac{1}{2}$ in the
definition of the (symmetrised) cuts. In practice it means that we take the
average of the two contributions in the $s$- and $t$-cuts, and multiply the
$u$-cut with a symmetry factor as two identical (super)particles cross the
cut.}
\begin{equation}
   \tilde{\cA}^{(1)}  (\bar{1}, 2, \bar{3}, 4 ) |_{s-\text{cut}}\ = \ \frac{1}{2}\int \! d^3\eta_{\ell_1} d^3\eta_{\ell_2} \ \tilde\cA^{(0)} (\bar1, 2, - \bar{\ell}_2, - \ell_1) \times \tilde\cA^{(0)} (\bar3, 4,  \bar{\ell}_1, \ell_2) + \ell_1 \leftrightarrow \ell_2 \,.
\end{equation}
The one-loop amplitude has cuts in the $s$-, $t$- and $u$-channels, for
which we find the following integrands:
\beqa
\label{allcuts}
  \tilde{\cA}^{(1)}  (\bar{1}, 2, \bar{3}, 4 ) |_{s-\text{cut}}
  & =  &
  {i \over 2}\, \cA^{(0) }  (\bar{1}, 2, \bar{3}, 4 )\ c_{s} \,\cS_{12} I (1,2,3,4) |_{s-\text{cut}} \ ,
   \nonumber \\
  \tilde{\cA}^{(1)} (\bar{1}, 2, \bar{3}, 4 ) |_{t-\text{cut}}
  & =  &
{i \over 2}\, \cA^{(0) }  (\bar{1}, 2, \bar{3}, 4 )\ c_{t} \, \cS_{23} I (1,2,3,4) |_{t-\text{cut}} \ ,
    \nonumber\\ 
 \tilde{\cA}^{(1)} (\bar{1}, 2, \bar{3}, 4 ) |_{u-\text{cut}}
 & =  &
 {i \over 2} \, \cA^{(0) }  (\bar{1}, 2, \bar{3}, 4 ) \   c_{u} \, \cS_{13} I (3,1,2,4) |_{u-\text{cut}} \ ,
\eeqa
where the colour factors $c_{s}$, $c_{t}$, $c_{u}$ are
\begin{align}\label{colourst}
    c_{s} & = N [1,2,3,4] + N [1,4,3,2] - 2 [1,2] [3,4]\, ,   \nonumber \\
    c_{t} & = N [1,2,3,4] + N [1,4,3,2] - 2 [1,4][3,2] \, , \nonumber \\
    c_{u} & = 2 [1,2] [3,4] - 2 [1,4][3,2]
\, ,
  \end{align}
and we recall that by $\cA^{(0) }  (\bar{1}, 2, \bar{3}, 4 ) $ we denote
the colour-ordered four-point amplitude.  Furthermore, we indicate by $
\cS_{ab} I (a,b,c,d) |_{s_{ab}-\text{cut}}$, the $s_{ab}$-cut of the
one-loop box function $I(a,b,c,d)$ in \eqref{YTbox}, symmetrised in the
cut loop momenta $\ell_1$ and $\ell_2$, which are defined such that
$\ell_1 + \ell_2 = p_a + p_b$,
  \begin{align}
   \cS_{12}I  (1,2,3,4) |_{s-\text{cut}}
    & =
    \frac{s \Tr (\ell_1 p_1 p_4)}{(\ell_1 - p_1)^2 (\ell_1 + p_4)^2} + \ell_1
    \leftrightarrow \ell_2 \, ,  \nonumber \\
  \cS_{23} I (1,2,3,4) |_{t-\text{cut}}  & =
    \frac{(-t)\Tr (\ell_1 p_1 p_2)}{(\ell_1 - p_1)^2 (\ell_1 + p_2)^2} + \ell_1
    \leftrightarrow \ell_2 \, , \nonumber \\
    \cS_{13} I(3,1,2,4) |_{u-\text{cut}}  & =
    \frac{u \Tr (\ell_2 p_3 p_4)}{(\ell_2- p_3)^2 (\ell_2 + p_4)^2} + \ell_1
    \leftrightarrow \ell_2 \, .\
     \end{align}
We should stress here that despite the simplified notation the cut momenta
$\ell_1$ and $\ell_2$ are different for the three distinct channels under
considerations. For instance,  $\ell_1 + \ell_2 = p_1 + p_2$ for the
$s$-cut, while $\ell_1 + \ell_2 = p_2 + p_3$ in the $t$-cut and $\ell_1 +
\ell_2 = p_1 + p_3$  in the $u$-cut.  Recall that the symmetrisation in the
cut momenta  in the $s$- and $t$-channel coefficients   originates from
summing over all possible particle species that can propagate on the cut
legs, while in the $u$ cut there is a single configuration allowed, and the
result turns out to be automatically  symmetric in  $\ell_1$ and $\ell_2$.

 Next we merge  the cuts into box functions. For the planar structures
 $[1,2,3,4]$ and $[1,4,3,2]$ this is immediate as the only function
 consistent with the $s$- and $t$-cuts in  \eqref{allcuts} and vanishing
 $u$-cut is $I(1,2,3,4)$.  Hence, the corresponding planar amplitude is
  \beq
  i\, \cA^{(0) }  (\bar{1}, 2, \bar{3}, 4 ) \  N \big([1,2,3,4] + [1,4,3,2] \big) \,  I(1,2,3,4)
  \ ,
  \eeq
thus arriving
at the expression \eqref{xxxxx} for the planar part of the full one-loop
amplitude.%
\footnote{Note that at the level of the integral we can simply
    replace  $ \cS_{12} I(1,2,3,4) $ by $ 2\, I(1,2,3,4) $. }
For the non-planar terms $[1,2][3,4]$ and $[1,4][3,2]$ we need  to use the
results of Appendix \ref{sub:interchangel1l2} and in particular
\eqref{symm33}, which we reproduce here,
\beq
\cS_{ab} I(a,b,c,d) |_{s_{ab}-\text{cut}} \ = \ \cS_{ab} I( a,b,d,c) |_{s_{ab}-\text{cut}}
\ .
\eeq
Firstly, we note that an immediate consequence of this result is that
  \beq
   \cS_{23} I  (2,3,4,1) |_{t-\text{cut}} \, - \,      \cS_{23} I (2,3,1,4) |_{t-\text{cut}} \ = \ 0\, ,
\,
\eeq
in other words the combination $I(2,3,4,1) - I(2,3,1,4)$, symmetrised in
the loop momenta $\ell_1$ and $\ell_2$, with $\ell_1+\ell_2 = p_2 + p_3$,
has a vanishing $t$-channel cut as expected for the coefficient of the
$[1,2][3,4]$ colour structure (see \eqref{colourst}). For the same
combination we find, using $I(2,3,4,1)=-I(1,2,3,4)$, the symmetrised
$s$-cut 
\beq
 - \cS_{12}   I(1,2,3,4) |_{s-\text{cut}}
\, ,
\eeq
and similarly, for the symmetrised $u$-cut we obtain
\beq
\cS_{13}   I(3,1,4,2) |_{u-\text{cut}}=   \cS_{13}   I(3,1,2,4) |_{u-\text{cut}}
  \, ,
\eeq
where we have used $I(2,3,1,4)=-I(3,1,4,2)$  and   \eqref{symm33}, which
allows us to swap the last two legs on the symmetrised $u$-cut.  Comparing
with  \eqref{allcuts} and \eqref{colourst} we can uniquely fix the
coefficient of the non-planar structure $[1,2][3,4]$:
  \beq
 2\,  i\, \cA^{(0) }  (\bar{1}, 2, \bar{3}, 4 ) \   [1,2]  [3,4] \ \Big[ I(2,3,4,1) - I(2,3,1,4)\Big]   \ ,
  \eeq
  or, using the first relation of \eqref{prima},
   \beq
 - 2\,  i\, \cA^{(0) }  (\bar{1}, 2, \bar{3}, 4 ) \   [1,2]  [3,4] \ \Big[ I(1,2,3,4) - I(4,2,3,1)\Big]   \ .
  \eeq
One can proceed similarly for the coefficient of the other non-planar
structure $[1,4][3,2]$, arriving  at the result quoted earlier in
\eqref{np1l}. Note that in that result we use the freedom to rename loop
momenta in order to eliminate the various symmetrisations introduced by
the operation $\cS_{ab}$ above.

\section{The Sudakov form factor at one and two loops}
\label{onebloop}

%%%%%%%%%%%%%%%%%%%%%%%%%%%%%%%%%%%%%%
We now move on to the form factors of gauge-invariant, single-trace scalar
operators
\begin{equation} \label{1} \mathcal{O} = \Tr \left (\phi^{A_{1}}
    \bar\phi_{B_{1}}
    \phi^{A_{2}}\bar\phi_{B_{2}}\dots\phi^{A_{L}}\bar\phi_{B_{L}} \right)
    \chi_{A_{1}\dots A_{L}}^{B_{1}\dots B_{L}}\,, \end{equation}
where $A$ and $B$ are indices of the $\mathbf{4}$ and   $\bar{\mathbf{4}}$
representation of the $R$-symmetry group $SU(4)$.  The operators \eqref{1}
are half BPS if $\chi$ is a symmetric traceless tensor  in all the $A_{i}$
and  $B_{i}$ indices separately  (see for example
\cite{Minahan:2008hf,Bak:2008cp}).  For $L=2$, the relevant operator is%
\footnote{More details on half-BPS operators, as well as conventions are
    discussed in Appendix \ref{sec:BPSdetails}. }
\begin{equation}
	\mathcal{O}^{A}{}_{B}
    =
    \Tr
    \left(
    \phi^{A} \bar\phi_B -
    \frac{\delta^{A}{}_{B}} {4}
    \phi^{K} \bar\phi_{K}
    \right) \, .
\end{equation}
In the rest of the paper we will focus on the Sudakov form factor
\beq
\label{chis}
\lan (\bar{\phi}_A )^{\bar{i}_1}_{i_1}  (p_1)\,
(\phi^4)^{i_2}_{\bar{i}_2}
(p_2) | \Tr (\bar\phi_A   \phi^4 ) (0)   | 0 \ran
\, := \,[1,2]\, \, F(q^2)
\ ,
\eeq
where $q := p_1 + p_2$ and $A \neq 4$, and we recall that $[1,2] :=
\delta^{\bar{i}_1}_{\bar{i}_2} \delta^{i_2}_{i_1}$.  At tree level,
\beq
\label{fftree}
F^{\rm (0)} (q^2) \ =\ 1
\ .
\eeq
We will now derive this quantity at one and two loops.

\subsection{One-loop form factor in ABJM}

\begin{figure}
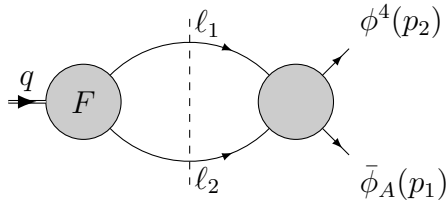

  		\centering
 		\qqcutoneloop
        \caption{\it The $q^2$ cut of the Sudakov form factor. Note that
        the amplitude on the right-hand side of the cut is summed over all
      possible colour orderings.} \label{fig:cutoneloop}
\end{figure}

At one loop it is possible to determine the integrand of the form
factor from a single unitarity cut in the $q^2$ channel. As shown in Figure
\ref{fig:cutoneloop}, on one side of the cut there is the Sudakov form
factor and on the other side the  complete four-point amplitude, both at
tree level.
The colour-ordered tree amplitude is given in
\eqref{eq:amptree}. Let us work out the colour factor first.
It is given by
\beq
\delta^{\bar{i}_{\ell_2}}_{\bar{i}_{\ell_1}} \delta^{i_{\ell_1}}_{i_{\ell_2}}
(\delta^{\bar{i}_1}_{\bar{i}_2}\delta^{i_2}_{i_{\ell_1}}
\delta^{\bar{i}_{\ell_1}}_{\bar{i}_{\ell_2}}\delta^{i_{\ell_2}}_{i_{1}}  -
\delta^{\bar{i}_{1}}_{\bar{i}_{\ell_2}}\delta^{i_{2}}_{i_{1}}
\delta^{\bar{i}_{\ell_1}}_{\bar{i}_{2}}\delta^{i_{\ell_2}}_{i_{\ell_1}} )\
= \ (N^\prime - N) \delta^{\bar{i}_1}_{\bar{i}_2}\delta^{i_2}_{i_{1}} \ .
\eeq
Obviously, the one-loop form factor vanishes identically in ABJM theory,
because in this case $N^\prime = N$.

We now consider the kinematic part.  Since the operator is built solely
out of scalars, only the four-point scalar amplitude can appear in the cut.
To match the particles of the tree amplitude in Figure
\ref{fig:cutoneloop},
we pick the $(\eta_1)^1(\eta_{\ell_1})^3(\eta_{\ell_2})^2(\eta_2)^0$
component from the $\delta^{6}(Q)$ to write the $q^2$ cut of the one-loop
form factor as:
\begin{equation}\label{thecomponent}
  \frac{\delta^{(6)}(Q)\bigr|_{(\eta_1)^1(\eta_{\ell_1})^3(\eta_{\ell_2})^2(\eta_2)^0}}
  {\ang{1\, 2}\ang{2\,\ell_1}}
  =
  \frac{\ang{\ell_1\,\ell_2}^2\ang{1\,\ell_1}}
  {\ang{1\,  2} \ang{2\,\ell_1}} \, = \,
  \frac{\ang{1\,2}\ang{1\,\ell_1}}
  {\ang{2\,\ell_1}} \, = \,
  -{{\rm Tr} (\ell_1 p_1 p_2)
\over 2 (  \ell_1 \cdot p_2)}
  \, ,
\end{equation}
which can be immediately lifted to a full integral as it is the
only possible cut of the form factor. Thus we get,
\begin{equation}
  \label{eq:OneLoopFF}
	F^{(1)}(q^2)
	=
	 (N^\prime - N) \,  \int\!\!
         \frac{d^D \ell_1} {i \pi^{D/2}}
         \frac{{\rm Tr} (\ell_1 p_1 p_2)}
         {\ell_1^2 \, (\ell_1 - p_2)^2 (\ell_1 - p_1-p_2)^2}\, .
\end{equation}
The integral in \eqref{eq:OneLoopFF} is a linear triangle and is of
$\mathcal{O}(\eps)$. Hence, we conclude that the one-loop Sudakov form
factor in ABJ theory vanishes in strictly three dimensions. Moreover, the
three-dimensional integrand vanishes in ABJM theory but is non-vanishing
for $N\neq N^\prime$ and  can (and does) participate in unitarity cuts at
two loops in ABJ theory.  
Note, that the vanishing of the one-loop form factors in ABJ(M)
is consistent with the infrared finiteness of one-loop amplitudes in
ABJ(M).

\subsection{Two-loop form factor in ABJM}

Next, we come to the computation of the two-loop Sudakov form factor.
In order to construct an ansatz for its integrand we will make use of
two-particle cuts, and fix potential remaining ambiguities with various
three-particle cuts described in detail in Sections \ref{sec:Three-vertex
cuts} and \ref{sec:three-particle-cut}.

Three-particle cuts are very useful  because they receive contributions
from planar as well as non-planar integral functions at the same time, and
thus are particularly constraining.  A special feature of ABJM theory is
that all amplitudes with an odd number of external particles vanish and, as
a consequence, all cuts involving such amplitudes are identically zero
\cite{Huang:2012hr}. In our case this observation will be important for
triple cuts, where three- and five-particle amplitudes would appear.

A particular type of such cuts, first considered in \cite{Huang:2012hr} in
the context of loop amplitudes in ABJM, involves three adjacent cut loop
momenta meeting at a three-point vertex.  The vanishing of these cuts
imposes strong constraints on the form of the loop integrands. We will
discuss and exploit this later in this section,  where we will also make
the intriguing observation that  integral functions with numerators
satisfying such constraints are transcendental and free of certain 
unwanted infrared divergences.

\subsubsection{Two-particle cuts}

\begin{figure}[h]
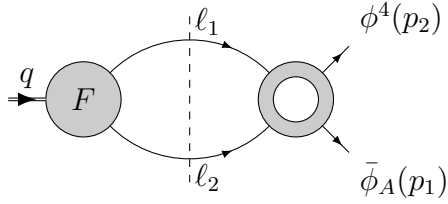

 \centering
    {\qqcuttwoloops}
     	\caption{\it Tree-level form factor glued to the complete one-loop amplitude.
  }
  \label{fig:cuttwoloops}
\end{figure}

We begin by considering the cut shown in Figure \ref{fig:cuttwoloops},
which contains a  tree-level Sudakov form factor merged with the integrand
of the complete one-loop, four-point  amplitude. The internal particle
assignment is fixed and is determined by the  particular operator we
consider. The integrand of this cut is  schematically given by
\beq
  F^{(0)} (\bar{\ell}_2, \ell_1) [\ell_2,\ell_1] \, \, \tilde\cA^{(1)}  \big( \bar\phi_A(p_1),   \phi^4(p_2), \bar\phi_4(-\ell_1), \phi^A(-\ell_2)
\big)
%\ + \ \ell_1 \leftrightarrow \ell_2
\ ,
\eeq
where we picked the relevant component amplitude of the complete one-loop amplitude $\tilde\cA^{(1)}$, given  in
\eqref{res1l}, and we recall that the colour factor $[a, b]$
is defined in \eqref{nota}.

We begin by  working out the colour structures that will appear in the
result. Firstly we consider the planar amplitude \eqref{xxxxx} and combine
it with the part of the non-planar amplitude \eqref{np1l} containing
$I(1,2,-\ell_1,-\ell_2)$.  Intriguingly, by contracting this with the
tree-level form factor (given in \eqref{chis} and \eqref{fftree})  we
obtain a vanishing result:
\beq
\label{cfp1}
\Big( N \big( [1,2,\ell_1, \ell_2] + [1,\ell_2,\ell_1,2] \big) - 2 [1,2] [\ell_1, \ell_2] \Big) [\ell_2,\ell_1]\  = \  0 \ .
\eeq
We now consider the remaining contributions arising from the non-planar
one-loop amplitude \eqref{np1l}.  There are two possible colour
contractions to consider,
\beq
\label{cfnp1}
c_{\rm NP}^{(1)} := 2\,  [1,2][\ell_1,  \ell_2] [\ell_2 , \ell_1]
 \ = \ 2\,  N^2 [1,2]
 \ ,
 \eeq
and
\beq
\label{subl1}
c_{\rm NP}^{(2)} :=2 \, [\ell_1,2][1,\ell_2] [\ell_2 , \ell_1]
 \ = \ 2\,   [1,2]
 \ .
 \eeq
Note that \eqref{subl1} is subleading in the large $N$  limit,
and can be discarded in the large-$N$ limit. Moreover, the corresponding
coefficient  actually vanishes which implies that the two-loop form factor
does not have non-planar corrections.

\begin{figure}[H]
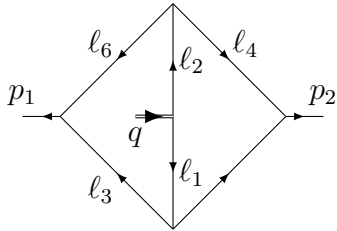

  \centering
  \nonplanar
  \caption{\it The crossed triangle integral arising from gluing  a tree form factor with the complete one-loop four-point amplitude.  The arrow in the middle denotes the location where the momentum
    $q=p_1+p_2$ is injected.  We call these integrals ``crossed triangles''
    because they have the topology of the master integral
    \eqref{eq:trianx}. Note however that the latter integral is non-transcendental, while the particular numerator in \eqref{eq:intXT} makes this integral transcendental.}
    \label{fig:nonplanar}
\end{figure}

We now need to determine the coefficient of $c_{\rm NP}^{(1)}$.
On the two-particle cut $\ell_1^2=\ell_2^2=0$ its integrand is given by the
appropriate component tree-level amplitude \eqref{thecomponent} times a
particular box integral \eqref{np1l}:
\beq
\label{xtint0}
\cC_1^{\rm (NP)} |_{s-\text{cut}} \, := \, {1\over 2} \frac{\langle 1 2 \rangle \langle 1 \ell_1 \rangle}{\langle 2 \ell_1 \rangle} \,
I(-\ell_2, 2, -\ell_1, 1) + \ell_1 \leftrightarrow \ell_2 \ .
\eeq
%%%%%%%%%%%%%%%%%%%%%%%%%%%%%%%%%%%%%%%
%
%
Recall that we have to symmetrise in order to include all particle species
in the sum over intermediate on-shell states. Since $I(-\ell_2, 2, -\ell_1,
1)$ is antisymmetric under $\ell_1 \leftrightarrow \ell_2$ the complete
cut-integrand can be written as\footnote{Similarly as done earlier for the complete
one-loop amplitude, we include a factor of $1/2$ in the symmetrisation.}
\beqa
\label{xtint}
\cC_1^{\rm (NP)} |_{s-\text{cut}}
&:=&
\frac{1}{2}
\left(
    \frac{\langle 1 2 \rangle \langle 1 \ell_1 \rangle}{\langle 2 \ell_1 \rangle}
    -
    \frac{\langle 1 2 \rangle \langle 1 \ell_2 \rangle}{\langle 2 \ell_2 \rangle}
\right)
\, I(-\ell_2, 2, -\ell_1, 1)
\\ \nonumber
&=&
-\frac{1}{2}\,  \int\!\frac{d^D \ell_3}
    {i \pi^{D/2} }\,
\frac{ q^2
   \big[ \Tr \left( p_1 p_2 \ell_1 \ell_3 \right) - q^2\ell_3^2\big] }
  { \ell_3^2 \, (\ell_1 - \ell_3)^2
  (p_1 - \ell_3)^2 (\ell_3 - \ell_1 + p_2)^2}\
\ .
\eeqa
Summarising, the  two-particle cut indicates that the two-loop form factor is
expressed in terms of a single crossed triangle with a particular
numerator, represented in Figure \ref{fig:nonplanar}, 
\begin{equation}
    \label{eq:intXT}
    \text{\bf XT}(q^2)
    \ = \ 
    \int\!
    \frac{d^D \ell_1 d^D \ell_3}
    {(i \pi^{D/2} )^2}
    \frac{
      q^2 \big[\Tr
      \left(
      p_1 p_2 \ell_1 \ell_3
      \right)
      - q^2 \ell_3^2 \big]
   }
    {\ell_1^2 \, \ell_2^2 \, \ell_3^2 \,
    (\ell_1 - \ell_3)^2
    (p_1 - \ell_3)^2 (\ell_3 - \ell_1 + p_2)^2} \, ,
\end{equation}
so that
\beq
\cC_1^{\rm (NP)} = -\frac{1}{2} \text{\bf XT}(q^2) \, .
\eeq
For future  convenience we will define
\begin{equation}
\label{xt-small}
       \text{\bf xt}
    \ := \
       \frac{
   q^2 \big[    \Tr
      \left(
      p_1 p_2 \ell_1 \ell_3
      \right)
      - q^2 \ell_3^2\big]
   }
    {\ell_1^2 \, \ell_2^2 \, \ell_3^2 \,
    (\ell_1 - \ell_3)^2
    (p_1 - \ell_3)^2 (\ell_3 - \ell_1 + p_2)^2} \, .
\end{equation}
The result of the evaluation of $\text{\bf XT}(q^2)$ is quoted in \eqref{final}
and   shows that this quantity has maximal  degree of transcendentality.
Before evaluating $\text{\bf XT}(q^2)$, we use triple cuts in order to
confirm the correctness of the ansatz obtained from two-particle cuts.

\subsubsection{Three-vertex cuts}
\label{sec:Three-vertex cuts}

To confirm the uplift of the two-particle cut to the integral
\eqref{eq:intXT}, we will study  additional cuts.  We begin by considering
three-point vertex cuts involving three adjacent legs meeting at a
three-point vertex. These cuts were first examined in \cite{Huang:2012hr},
where it was observed that they must vanish since  there are no
three-particle amplitudes in ABJM theory.  Calling $k_1$, $k_2$ and $k_3$
the momenta meeting at the vertex, we have
\begin{equation}
  \label{eq:trivertex}
  k_1 + k_2 + k_3 = 0 \, ,
  \qquad
  k_1^2   = k_2^2 = k_3^2 = 0 \, .
\end{equation}
The conditions \eqref{eq:trivertex} imply that all
spinors associated to these  momenta are proportional, thus
\begin{equation}
  \ang{k_1 \, k_2}
  =
  \ang{k_2 \, k_3}
  =
  \ang{k_3 \, k_1}
  = 0 \, .
\end{equation}
As an example consider the three-point vertex
cut of $\text{\bf XT}(q^2)$ with momenta $\ell_2$, $\ell_4$ and $\ell_6 := \ell_2 - \ell_4$
(see Figure \ref{fig:nonplanar} for the labelling of the momenta).
 Importantly, the form factor is expected to vanish as the three momenta
belonging to a three-point vertex become null.
By rewriting the numerator of \eqref{eq:intXT}  using only cut momenta, it is immediately seen 
that it vanishes, since 
\beqa
\label{sso}
    &&\Tr \big[p_1 p_2 (p_1-\ell_2) (p_1-\ell_6 )\big] - q^2  (p_1 - \ell_6)^2   \ = \ 
    -\Tr \big[p_1 p_2 (p_1-\ell_2) \ell_6 \big] - q^2  (p_1 - \ell_6)^2  \nonumber \\
   &&=\ -\Tr (p_1 p_2 p_1 \ell_6 ) + 4  (p_1 \cdot p_2)  (p_1 \cdot \ell_6)\ =  \ 0 \ ,
  \eeqa
where we have used $ \langle \ell_2 \ell_6 \rangle =0$ to set $\Tr (p_1 p_2 \ell_2\ell_6 )=0$.  It is easy to see that all
other three-vertex cuts of the integral \eqref{eq:intXT} vanish in a
similar fashion because of the particular form of its numerator.

Important consequences of these specific properties of the numerator of the
integral function \eqref{eq:intXT} are that the result is transcendental as
we will show below and is free of unphysical infrared divergences related
to internal three-point vertices. These divergences appear generically in
three-dimensional two-loop integrals with internal three-vertices even if the
external kinematics is massive (unlike in four dimensions) and it appears
that master integrals with appropriate numerators to cancel these peculiar
infrared divergences are a preferred basis for amplitudes  and form factors in
ABJM.  Related discussions in the context of ABJM amplitudes have appeared
in \cite{Huang:2012hr, Bianchi:2011fc}. Note that for form factors we do
not have dual conformal symmetry, which gives further constraints on the structure of the numerators of integral functions appearing in amplitudes.

\subsubsection{Three-particle cuts}
\label{sec:three-particle-cut}

\begin{figure}[H]
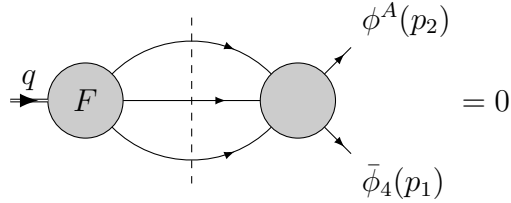

    \begin{center}
    \qqtriplecut
  \end{center}
  \caption{\it The (vanishing) three-particle cut of the two-loop Sudakov form factor.}
  \label{fig:triplecutplanar}
\end{figure}
\noindent
The remaining cut we will study is a triple cut of the type illustrated in Figure
\ref{fig:triplecutplanar}. These cuts may potentially detect additional
integral functions which have no two-particle cuts at all, and are thus very
important.  Moreover, such cuts are sensitive to both planar and non-planar
topologies.  In this triple cut,  a  tree-level amplitude is connected to a
tree-level form factor by three cut propagators. Due to the vanishing of
amplitudes with an odd number of external legs in the ABJM theory, the
triple cut in question vanishes.  We will now check  that the triple cut of
the two-loop crossed triangle  {\bf XT} of   \eqref{eq:intXT}, which we
have detected using  two-particle cuts, is indeed equal to zero. %

\begin{figure}
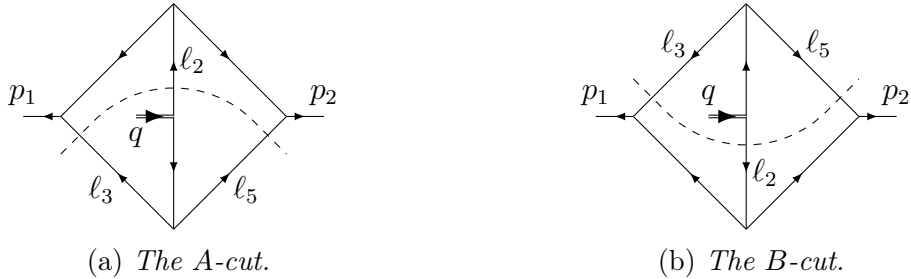

\begin{centering}
  \begin{subfigure}[b]{0.45\textwidth}
    \centering
    {\nonplanarAcut}
    \caption{\it The $A$-cut.}
    \label{fig:nonplanarA}
  \end{subfigure}%
  \quad
  \begin{subfigure}[b]{0.45\textwidth}
    \centering
    {\nonplanarBcut}
    \caption{\it The $B$-cut.}
    \label{fig:nonplanarB}
  \end{subfigure}%

\par\end{centering}

\caption{\it The two triple cuts of the crossed triangle, with $\ell_2 + \ell_3 + \ell_5 = q$. In the second figure we have relabelled the loop momenta in order to merge the two contributions. }

\end{figure}

To this end, we note that there are two possible ways to perform a triple-cut on {\bf
XT},  shown in Figures \ref{fig:nonplanarA} and \ref{fig:nonplanarB}.  The
cut loop momenta are called $\ell_2$, $\ell_5$ and $\ell_3$ and  satisfy
\begin{equation}
  \label{eq:momconstriple}
  \ell_2 + \ell_5 + \ell_3
  =
  p_1 + p_2 \,,
  \qquad
  \ell_2^2 = \ell_5^2 = \ell_3^2 = 0 \,.
\end{equation}
We observe  that these two cuts cannot be converted into one another by a
simple relabelling of the cut momenta because of the  
non-trivial numerators.
The A-cut  depicted in Figure \ref{fig:nonplanarA}
of the non-planar integrand is:
\begin{equation}
  \label{eq:nonpA}
\text{\bf XT}\bigr|_{\text{3-p cut A}}
  =
  -q^2\frac{\ang{1\,2}}
  {\ang{\ell_3\,\ell_5}\ang{\ell_5 \, 2 }\ang{1\, \ell_3}}
  \ .
\end{equation}
\noindent
After a similar calculation, the $B$-cut of this integral, depicted in
Figure \ref{fig:nonplanarB}, turns out to be identical to the $A$-cut:
\begin{equation}
  \text{\bf XT}\bigr|_{\text{3-p cut B}}
  =
  \text{\bf XT}\bigr|_{\text{3-p cut A}}
  =
 -q^2\frac{\ang{1\,2}}
  {\ang{\ell_3\,\ell_5}\ang{\ell_5 \, 2 }\ang{1\, \ell_3}}
  \ .
\end{equation}
A quick way to establish the vanishing of the triple cuts consists in
symmetrising in the particle momenta $p_1$ and $p_2$, which is allowed
since the Sudakov form factor is a function of $q^2$. This symmetrisation
gives
\beq
- {q^2\ang{1\,2}\over \ang{\ell_3\,\ell_5} }
 \left[ \frac{1}
  {\ang{\ell_5\, 2}\ang{1\, \ell_3}} - \frac{1}
  {\ang{\ell_5\,1}\ang{2\, \ell_3}} \right]
  \ = \
  -\,  {q^4\over    \langle1 | \ell_5| 2\rangle \,  \langle1 | \ell_3 |2\rangle}
    \ .
\eeq
This expression is symmetric in $\ell_5$ and $\ell_3$. In evaluating the
triple cut one has to introduce a jabobian proportional to $\eps(\ell_2,
\ell_3, \ell_5)$ \cite{Huang:2012hr} which effectively makes this triple
cut vanish upon integration. This implies that the complete answer for the two-loop form factor in ABJM is proportional $\text{\bf XT}(q^2)$ and no additional integral
functions have to be introduced.

%%%%%%%%%%%%%%%%%%%%%%%%%%%%%%%%%%%%
\subsubsection{Results and comparison to the two-loop amplitudes}

Combining the information from  the unitarity cuts discussed above, we
conclude that the two-loop Sudakov form factor in ABJM is given by a single
non-planar integral
\begin{equation}
  \label{eq:resultsymb}
  F_\text{ABJM}(q^2)
\   =\
  -2\, \left(\frac{N}{k}\right)^2\, \left(-\frac{1}{2}\right){\bf XT}(q^2)  \, ,
\end{equation}
where  ${\bf XT}(q^2)$ is  defined in \eqref{eq:intXT} and we have
reintroduced the dependence on the Chern-Simons level $k$.  The integral
${\bf XT}(q^2)$ can be computed by reduction to master integrals using
integration by parts identities. The details of the reductions are provided
in Appendix \ref{sec:evaluating-integrals}. The expansion of the result in
the dimensional regularisation parameter $\eps$ can then be found  using
the expressions for the  the master integrals
\eqref{eq:sunset}--\eqref{eq:trianx}. Plugging these masters into
the reduction \eqref{eq:redxt}, we arrive at
\begin{align}
\label{final}
  \text{\bf XT}(q^2)
  &
  =\left(\frac{-q^2 e^{ \gamma_E} }{\mu^2}\right)^{-2\epsilon}\biggl[
  \frac{\pi}{\epsilon^2}
  + \frac{2\pi\log{2}}{\epsilon}
  - 4\pi\log^2{2}
  - \frac{2\pi^3}{3}
  + \cO(\epsilon)\biggr]\, ,
  \end{align}
where $\gamma_E$ is the Euler-Mascheroni constant.
One comment is in order here. We have derived \eqref{final} in a
normalisation where the the loop integration measure is written as $d^Dl /
(i \pi^{D/2})$. This should be converted to the standard one $d^Dl /
(2\pi)^D$. At two loops, this implies that  \eqref{final} has to
be multiplied by  a factor of $- 1/ (4 \pi)^D$. The result in the  standard
normalisation is then
 \begin{equation}
\label{final2}
  \cF_\text{ABJM}(q^2)\ = \    -{1\over  (4 \pi)^{3} }\left({N\over k}\right)^2
   \left(\frac{-q^2 e^{ \gamma_E} }{4 \pi \mu^2}\right)^{-2\epsilon}
\begin{aligned}[t]
    &
\biggl[
  \frac{\pi}{\epsilon^2}
  + \frac{2\pi\log{2}}{\epsilon}
  - 4\pi\log^2{2}
  - \frac{2\pi^3}{3}
  + \cO(\epsilon)\biggr]\, .
  \end{aligned}
\end{equation}
We note that  $ \cF(q^2)$  can be expressed more compactly  by introducing a new scale
\beq
\label{mup}
{\mu^\prime}^2  := 8 \pi \, e^{- \gamma_E}\mu^2
\ ,
\eeq
in terms of which we  get
\begin{equation}
\label{final3}
  \cF_\text{ABJM}(q^2)\ = \    {1\over  64\pi^{2}} \left( \frac{N}{k}\right)^2
   \left(\frac{-q^2  }{ {\mu^\prime}^2}\right)^{-2\epsilon}\biggl[
  -\frac{1}{ \epsilon^2}\, +\,  6 \log^2 2 \, +\, {2 \pi^2 \over 3}  \, +\, \cO(\epsilon) \biggl]
  \, ,
\end{equation}
which is our final result.

We now discuss two consistency checks that confirm the correctness of
\eqref{final3}.  Firstly, we recall that  the Sudakov form factor captures
the infrared divergences of scattering amplitudes. We now check that
\eqref{final3} matches the infrared poles of the four-point amplitude
evaluated in \cite{Chen:2011vv, Bianchi:2011dg}.  Here we quote its
expression    as given in \cite{Bianchi:2011dg}:
\beq
\label{silvia}
\cA^{\rm (2)}_4
\ = \ - {1\over 16 \pi^2} \cA^{(0)}_4 \left[
 { ( - s / {\mu^\prime}^2 )^{- 2 \eps} \over 4 \eps^2}
 +
  { ( - t / {\mu^\prime}^2 )^{- 2 \eps} \over 4 \eps^2} - {1\over 2} \log^2 \left( {-s \over -t} \right) - 4\zeta_2- 3 \log^2 2 \right]
\ ,
\eeq
where $\mu^\prime$ is related to $\mu$ in the same way as in \eqref{mup}.
Hence, the Sudakov form factor \eqref{final3} is in perfect agreement with
the form of the infrared divergences of \eqref{silvia}.  Secondly, we have
also checked that the expansion of our result in terms of master integrals
(i.e.~the expansion of the two-loop non-planar triangle {\bf XT} defined in
 \eqref{eq:intXT}) is identical to that obtained from the Feynman diagram
 based result of \cite{donovan}. This implies that the cut-based
 calculation of this paper and the Feynman diagram calculation of
 \cite{donovan} agree to all orders in $\eps$ -- even though we have been using
 cuts in strictly three dimensions.

\section{Maximally transcendental integrals in 3d} % (fold)
\label{bonus}
As discussed in section \ref{sec:Three-vertex cuts}, the integrand
{\bf xt} that appears in the Sudakov form factor in ABJM has a particular
numerator such that all the cuts which isolate a three-point vertex
vanish. We have observed in this example that this property ensures that
the integral {\bf XT}  has a uniform (and maximal) degree of
transcendentality -- failure to obey the triple-cut condition, for instance
by altering the form of the numerator, would result in the appearance of
new terms with lower degree of transcendentality.  In this section we
present further integrals that vanish in these three-particle cuts and have
maximal  degree of  transcendentality. These integrals are expected to
appear in the form factor of ABJ theory where cancellations between colour
factors such as that in \eqref{cfp1}, do not occur.

\begin{figure}[h]
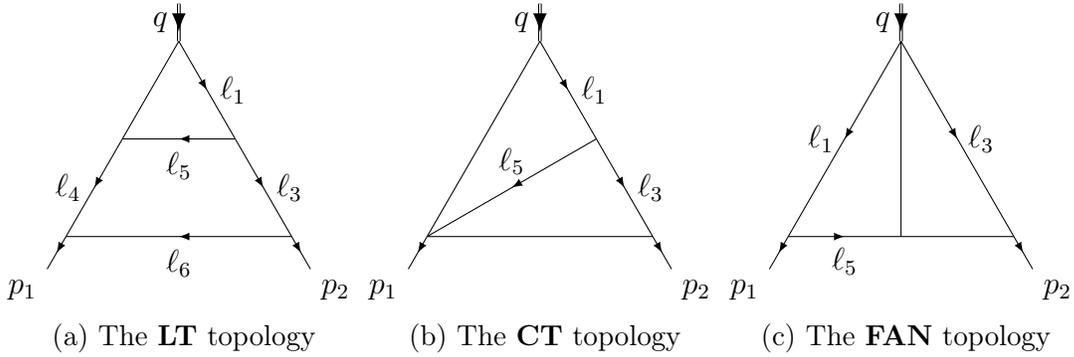

 \centering
  \begin{subfigure}[b]{0.3\textwidth}
    \centering
    {\laddertriangle}
    \caption{The {\bf LT} topology}
    \label{fig:LT}
  \end{subfigure}%
  \begin{subfigure}[b]{0.3\textwidth}
    \centering
    {\crooked}
    \caption{The {\bf CT} topology}
    \label{fig:CT}
  \end{subfigure}%
  \begin{subfigure}[b]{0.3\textwidth}
    \centering
    {\fan}
    \caption{The {\bf FAN} topology}
    \label{fig:FAN}
  \end{subfigure}%
        \caption{\it The three maximally transcendental integrals
            considered in  \eqref{eq:lt}, \eqref{aaaw} and \eqref{bbbw}.  }
  \label{fig:bonusintegrals}
\end{figure}

We begin by  considering the following planar integral function:
{
\setlength{\jot}{10pt}
  \begin{equation}
    \label{eq:lt}
    \text{\bf LT}(q^2)
    \begin{aligned}[t]
      &=\int\!\! \frac{d^D \ell_1 d^D \ell_3}{(i \pi^{D/2} )^2}
      \frac{- q^2 \left[ \, \Tr (p_1\,\ell_3\, p_2\, \ell_1) - (\ell_1
          - p_1)^2 (\ell_3 - p_2)^2 \right]}
      {\ell_1^2 \, (p_1 + p_2 - \ell_1)^2 \,\ell_3^2 \, (p_1+p_2 - \ell_3)^2 (\ell_1 - \ell_3)^2 (\ell_3 - p_2)^2}\,\\
      &=
      \left(\frac{-q^2e^{\gamma_E}}{\mu^2}\right)^{-2\epsilon}\biggl[-
      \frac{\pi}{4\epsilon^2} - \frac{\pi\log{2}}{\epsilon} +
      2\pi\log^2{2} - \frac{5\pi^3}{8} + \cO(\epsilon)\biggr]\, ,
    \end{aligned}
\end{equation}
}
which is shown in Figure \ref{fig:LT}.
It is easy to see that the three vertex cut $\{\ell_1, \ell_3, \ell_5\}$
vanishes, since on this cut the numerator can be placed in the form
\begin{equation}
  \label{eq:ansatznumeratoronvertexcut}
  \ang{\ell_1 \, 1} \ang{\ell_3 \, 2} \ang{1 \, 2} \ang{\ell_3 \, \ell_1} \, ,
\end{equation}
after using a Schouten identity.  \eqref{eq:ansatznumeratoronvertexcut}
vanishes because  $\ang{\ell_3 \, \ell_1}=0$  on this cut.

A further property of \eqref{eq:lt} emerges when we consider particular
triple cuts  involving two adjacent massless legs,  which in three
dimensions are associated with soft gluon exchange \cite{Huang:2012hr}.
With reference to Figure \ref{fig:LT}, we cut the three momenta $\ell_3$,
$\ell_6$ and $\ell_4$. The cut conditions $ \ell_3^2 = \ell_6^2 = \ell_4^2
=0$ together with the masslessness of  $p_1$ and $p_2$ can only be
satisfied if $\ell_6$ becomes  soft, that is
\begin{align}
    \ell_6   \to 0
    \,, \quad \ell_4  \to p_1\,, \quad \ell_3  \to p_2 \, .
    \label{eq:softlim}
\end{align}
 In this limit,  the second term of \eqref{eq:lt} vanishes  since $\ell_3 -
 p_2 =  \ell_6 \to 0$. The first term becomes
\begin{align}
  -  q^2 \frac{\Tr (p_1\,\ell_3\, p_2\, \ell_1)}
  {8 \epsilon (\ell_3 , p_1 , p_2)}
  \to
  - q^2 \frac{ \ang{ 2 | \ell_1 | 1 }}
  {4 \ang{1 2 }} \, ,
\end{align}
where  $8 \epsilon (\ell_3, p_1, p_2)$ is the Jacobian.%
\footnote{This Jacobian arises from re-writing the $\delta$-functions of the cut momenta,
$\ell_3^2 = \ell_4^2 = \ell_6^2 = 0 $,  in terms of $p_1, p_2$ and
$\ell_6$. }
After restoring the remaining propagators we are left with
\begin{align}
   \frac{2 \epsilon (\ell_1, p_1, p_2) }
         {\ell_1^2(\ell_1-p_2)^2(q-\ell_1)^2} \, ,
\end{align}
which reproduces  the one-loop integrand of the one-loop form factor,
given earlier in  \eqref{eq:OneLoopFF}.

Other examples of integrals with different topologies that satisfy the
three-particle cut condition are depicted in Figures \ref{fig:CT} and
\ref{fig:FAN}. The definitions of the integrals as well as their values are
listed below:
{ \setlength{\jot}{10pt}
\begin{align}
  \label{aaaw}
    \text{\bf CT}(q^2)
    &=
    \begin{aligned}[t]
      & \int\!\! \frac{d^D \ell_1 d^D \ell_3}{(i \pi^{D/2} )^2}
      \frac{\Tr(p_1,p_2,\ell_3,\ell_1)}
      {\ell_1^2 \, (p_1 + p_2 - \ell_1)^2 \,\ell_3^2 \,  (\ell_1 - \ell_3)^2\, (\ell_3 - p_2)^2}\,\\
      &= \left(\frac{-q^2 e^{ \gamma_E}
        }{\mu^2}\right)^{-2\epsilon}\biggl[ - \frac{\pi}{4\epsilon^2}
      + \frac{7\pi^3}{24} + \cO(\epsilon)\biggr]\, ,
    \end{aligned}\\ \cr
    \text{\bf FAN}(q^2)
    \label{bbbw}
    &=
    \begin{aligned}[t]
      & \int\!\! \frac{d^D \ell_1 d^D \ell_3}{(i \pi^{D/2} )^2}
      \frac{\Tr(p_1,p_2,\ell_3,\ell_1)}
      {\ell_1^2 \, \ell_3^2\,(p_1 + p_2 - \ell_1-\ell_3)^2 \,  (\ell_1 - p_1)^2\, (\ell_3 - p_2)^2}\\
      &= \left(\frac{-q^2 e^{ \gamma_E}
        }{\mu^2}\right)^{-2\epsilon}\biggl[ - \frac{\pi}{4\epsilon^2}
      + \frac{7\pi^3}{24} + \cO(\epsilon)\biggr]\, .
    \end{aligned}
\end{align}
}
\noindent
Note that the $\eps$ expansion of \eqref{aaaw} and \eqref{bbbw} agree up to
$\cO(1)$.  It is  simple to show that these integrals satisfy 
the  properties discussed earlier, for example by setting $\{\ell_1,\ell_3,\ell_5\}$
on shell in {\bf CT} and $\{\ell_1,p_1,\ell_5\}$ in {\bf FAN} and similarly
for all other possible three-vertex cuts.

The reductions of the integrals considered in this section in terms of
scalar master integrals through IBP identities can be found in Appendix
\ref{sec:evaluating-integrals}.

\section*{Acknowledgements}

It is a pleasure to thank  Marco Bianchi, Massimo Bianchi, Silvia Penati,
and especially Congkao Wen and Konstantin Wiegandt for stimulating  discussions, and
Donovan Young for informing us of his independent calculation of the
Sudakov form factor from Feynman diagrams \cite{donovan}.  \"OG and DK
would like to thank Jurgis Pa\u{s}ukonis for several \texttt{Mathematica}
consultations.  This work was supported by the STFC Grant ST/J000469/1,
``String theory, gauge theory \& duality''.

%%%%%%%%%%%%%%%%%%%%%%%%%%%%%%%%%%%%%%%%%%%%%%%%%%%%%%%
%%% Appendix starts here
%%%%%%%%%%%%%%%%%%%%%%%%%%%%%%%%%%%%%%%%%%%%%%%%%%%%%%%

\appendix

\section{Half-BPS operators}
\label{sec:BPSdetails}
In this appendix we briefly recall how half-BPS operators are introduced in ABJM theory.
Consider the variation of operators of the form $\Tr \left( \phi^{I} \bar\phi_{J} \right)$
with $I\neq J$. Setting for example $I = 1$ and $J = 4$, this expands to
\begin{equation}
	\delta \Tr
    \left(
      \phi^1\bar\phi_{4}
    \right)
	=
    \Tr
      \left(
        \delta \phi^1 \bar\phi_4 + \phi^1 \delta \bar\phi_4
      \right) .
\end{equation}
 Following \cite{Terashima:2008sy}, we use the transformations:
\begin{eqnarray}
	\delta \phi^{I}
	& = &
	i\,\omega^{IJ} \psi_{J}\, , \\
	\delta \bar{\phi}_{I}
	& = &
	i\,\bar{\psi}^{J} \omega_{IJ}\,.
\end{eqnarray}
The $\omega_{IJ}$'s are given in terms of the $(2+1)$-dimensional
Majorana spinors, $\epsilon_{i}$ $(i=1,\dots,6)$ which are the supersymmetry
generators:
\begin{eqnarray}
	\omega_{IJ} & = & \epsilon_{i}(\Gamma^{i})_{IJ}\, , \\
	\omega^{IJ} & = & \epsilon_{i}\left((\Gamma^{i})^{*}\right)^{IJ}\, ,
\end{eqnarray}
that are antisymmetric in $I,J$.  The $4\times4$ matrices $\Gamma^{i}$
are given by:
\begin{gather}
	\Gamma^{1}
    =
    \sigma_{2} \otimes 1_{2}\, ,
    \qquad
    \Gamma^{4}
    =
    -\sigma_{1} \otimes \sigma_{2}\, ,\\
	\Gamma^{2}
    =
    -i\sigma_{2} \otimes \sigma_{3}\, ,
    \qquad
    \Gamma^{5}
    =
    \sigma_{3} \otimes \sigma_{2}\,,\\
	\Gamma^{3}
    =
    i\sigma_{2} \otimes \sigma_{1}\,,
    \qquad
    \Gamma^{6}
    =
    -i 1_{2} \otimes \sigma_{2}\,,
\end{gather}
and satisfy the following relations,
\begin{eqnarray}
	\left\{ \Gamma^{i},\Gamma^{j\dagger}\right\}
    =
    2\delta_{ij}\,, & \left(\Gamma^{i}\right)_{IJ}
    =
    -\left(
      \Gamma^{i}
    \right)_{IJ}\, ,\\
	\frac{1}{2} \epsilon^{IJKL} \Gamma_{KL}^{i}
    = &
    -\left(
        \Gamma^{j\dagger}
      \right)^{IJ}
    =
    \left(
      \left(
        \Gamma^{i}
      \right)^{*}
    \right)^{IJ}\,,
\end{eqnarray}
 leading to
\begin{equation}
	\left(
      \omega^{IJ}
    \right)_{\alpha}
    =
    \left(
      \left(
        \omega_{IJ}
      \right)^{*}
    \right)_{\alpha} \, ,
    \qquad
    \omega^{IJ}
    =
    \frac{1} {2} \epsilon^{IJKL} \omega_{KL} \, .
  \label{eq:omegarelations}
\end{equation}
Explicitly, $\omega_{IJ}$ is given by the following matrix:
\begin{eqnarray}
	\omega_{IJ}                 &
    =                           &
    \left(\begin{array}{cccc}
	0                           & -i\epsilon_{5}-\epsilon_{6} & -i\epsilon_{1}-\epsilon_{2} & \epsilon_{3}+i\epsilon_{4}\\
	i\epsilon_{5}+\epsilon_{6}  & 0                           & \epsilon_{3}-i\epsilon_{4}  & -i\epsilon_{1}+\epsilon_{2}\\
	i\epsilon_{1}+\epsilon_{2}  & -\epsilon_{3}+i\epsilon_{4} & 0                           & i\epsilon_{5}-\epsilon_{6}\\
	-\epsilon_{3}-i\epsilon_{4} & i\epsilon_{1}-\epsilon_{2}  & -i\epsilon_{5}+\epsilon_{6} & 0
	\end{array}\right)\, .
\end{eqnarray}
 The term $\phi^1\delta\bar\phi_4$ yields
\begin{equation}
	\phi^1\delta\bar\phi_4
    =
    \phi^1\left[
      - \bar\psi^1 (\epsilon_3   + i \epsilon_4)
      + i \bar\psi^2 (\epsilon_1 + i \epsilon_2)
      - i \bar\psi^3 (\epsilon_5 + i \epsilon_6)
                                 + 0
      \right] .
\end{equation}
 Therefore, requiring $\phi^{1}\delta\bar{\phi}_{4}=0$ the conditions
are:
\begin{equation}
\label{eq:condomega}
  \begin{aligned}
    \epsilon_1 + i\epsilon_2 & = & 0\, , \\
    \epsilon_3 + i\epsilon_4 & = & 0\, ,  \\
    \epsilon_5 + i\epsilon_6 & = & 0\, ,
  \end{aligned}
\end{equation}
 which relate half of the generators with the other half
by constraining the components $\omega_{4J}=0$.

Note that because of the relations (\ref{eq:omegarelations}) which
set components of the form $\omega_{4L}$ to zero, the entries $\omega^{IJ}$
with $I,J\in(1,2,3)$ automatically vanish implying that
$\delta\phi^{I}=0\iff I\in(1,2,3)$.  This procedure may be iterated to show
that generally the operators $\Tr\left( \bar{\phi_{I}} \phi^{J} \right)$
for $I \neq J$ are indeed half-BPS. In the present work the operators under
consideration are of the type
\begin{equation}
  {\cal O}
  =
  \Tr\, (\phi^A \bar\phi_4) \, ,
\end{equation}
where $A\neq 4$.

% Section: BPS Operators (end)

\section{Properties of the box integral} % (fold)
\label{sec:box-prop}
\begin{figure}[H]
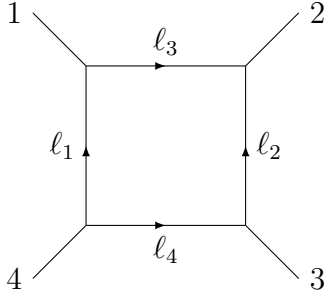

  \begin{centering}
    \begin{minipage}[c][1.3\totalheight]{1\columnwidth}%
      \begin{center}
         \fourbox{0.5}{$\ell_3$}{$\ell_2$}{$\ell_4$}{$\ell_1$}
      \par\end{center}%
    \end{minipage}
  \par\end{centering}
  \caption{\it Four-point one-loop box.\label{fig:4pt1loop}}
\end{figure}
The box integral function \eqref{YTbox}  was constructed and used in
\cite{Chen:2011vv}, and has several interesting properties that have been
exploited in the present work. This section presents and proves (some of)
these properties.

\subsection{Rotation by $90^\circ$ } % (fold)
\label{sub:rotateby90}

The first property we wish to  discuss is what could be called a
$\pi/2 $ rotation symmetry. Focusing on the numerator of the box
integrand,
\begin{equation}
\label{NN2}
    N \ = \  s \, \Tr (\ell_1 p_1 p_4) + \ell_1^2 \, \Tr (p_1 p_2 p_4) ,
\end{equation}
we can eliminate  $\ell_1$  in favour of
$\ell_3$ and arrange to  have only the external legs $p_2, p_3, p_1$ appear
in the numerator. Using momentum conservation, we can re-write  $N$ as
\begin{align}
\label{NNN}
    N  & =
    \left( -t -u \right)
    \Tr \left( (\ell_3 + 1 ) p_1 ( -p_1 -p_2 - p_3) \right)
     + \left( \ell_3 + p_1 \right)  \Tr\left( p_1 p_2 (-p_1 - p_2 - p_3) \right)\\ \nonumber\cr
    & =
    - \left[
          t \, \Tr (\ell_3 p_2 p_1) + \ell_3^2 \, \Tr(p_2 p_3 p_4)
      \right]
    + \mathcal{R} \, ,
\end{align}
where
\begin{equation}
    \mathcal{R}  = s \, \Tr (\ell_3 p_3 p_1) - u \, \Tr (\ell_3 p_2 p_1) - 2 (\ell_3 \cdot
    p_1)  \Tr (p_2 p_3 p_1) \, .
\end{equation}
In three dimensions the loop momentum $\ell_3$  can be expressed as a function of the
external momenta $p_1$, $p_2$, $p_3$ as
\begin{equation}
    \ell_3 = \alpha p_1 + \beta p_2 + \gamma p_3
    \, ,
\end{equation}
where $\alpha,\beta,\gamma$ are arbitrary coefficients. When this
identity is used in the expression for $\mathcal{R}$, we find that
$\mathcal{R}$ vanishes identically zero in  three dimensions.  Hence%
\begin{equation}
\label{nidentity}
    s \, \Tr (\ell_1 p_1 p_4) + \ell_1^2 \, \Tr (p_1 p_2 p_4)\  =\
          - t \,\Big(  \Tr (\ell_3 p_2 p_1) + \ell_3^2 \, \Tr(p_2 p_3 p_4) \Big) \, .
\end{equation}
It is also interesting to write down explicitly  the $s$- and $t$-cut of
the one-loop box.  Starting from the expression of the box integral
\begin{equation}
    \label{YTbox2}
I(1,2,3,4) \ := \   \int\!\!\frac{d^D \ell}{i \pi^{D/2}}\,
{ N\over \ell^2 (\ell - p_1)^2(\ell - p_1 - p_2)^2(\ell + p_4)^2} \, ,
\end{equation}
with $N$  given in \eqref{NN2},  we first consider the $s$-cut of this
function. This gives
\beq
  I(1,2,3,4)|_{s\text{-cut}}
  \  = \
  \frac{s \, \Tr (\ell_1 p_1 p_4)}{\ell_3^2 \, \ell_4^2} \, ,
  \eeq
which upon using $\ell_3 = \ell_1 - p_1$ and $\ell_4 = - (\ell_1 + p_4)$
becomes
\beq
I(1,2,3,4)|_{s\text{-cut}}
\  = \ \frac{s \langle 4 1 \rangle}{\langle 4 \ell_1 \rangle \langle \ell_1 1
\rangle} \ .
\label{iscut}
\eeq
Similarly the $t$-channel expression of the full integrand is
\begin{equation}
  I(1,2,3,4) =
  \frac{t \, \Tr (\ell_3 p_2 p_1) + \ell_3^2 \, \Tr(p_2 p_3 p_1)}
  {\ell_1^2 \, \ell_2^2 \, \ell_3^2 \, \ell_4^2}    \, .
\end{equation}
The $t$-cut of $I(1,2,3,4)$ is immediately written using the
three-dimensional identity \eqref{nidentity},
\begin{align}
  I(1,2,3,4)|_{t\text{-cut}}
  & =
    - \frac{t \, \Tr(\ell_3 p_2 p_1)}{\ell_1^2 \, \ell_2^2} \\ \nonumber
    & = \frac{t \langle 1 2 \rangle}{\langle 1 \ell_3 \rangle \langle
    \ell_3 2 \rangle} \, .
\end{align}
Finally, if we re-introduce the tree-level amplitude prefactor $
\mathcal{A}^{\text{(0)}}(\bar1, 2, \bar3, 4)   = 1 / ( \langle 1 2 \rangle
\langle 2 3 \rangle )$, we can write down the two cuts of the one-loop
amplitude,
\begin{align}
   \mathcal{A}^{\text{(0)}}(\bar1, 2, \bar3, 4)\times I(1,2,3,4) |_{s\text{-cut}}
  & =
      -\frac{\langle 3 4 \rangle}{\langle 4 \ell_1 \rangle \langle \ell_1 1
    \rangle} \, , \\
    \mathcal{A}^{\text{(0)}}(\bar1, 2, \bar3, 4) \times I(1,2,3,4) |_{t\text{-cut}}
  & =
      \frac{\langle 2  3 \rangle}{\langle 1 \ell_3 \rangle \langle \ell_3 2
    \rangle} \, .
\end{align}
%

% subsection Rotation by \frac{\pi}{2} (end)
\subsection{An identity for the $s$-channel cuts of $I(1,2,3,4)$ and $I(1,2,4,3)$} % (fold)
\label{sub:interchangel1l2}

Here we discuss an intriguing  property of the three-dimensional cuts of
$I(1,2,3,4)$.  We consider the $s$-channel cut of this function and
symmetrise it  in the cut loop momenta  $\ell_1$ and $\ell_2$, where
$\ell_1 + \ell_2 = p_1 + p_2$.
The result we wish to show is that the symmmetrised three-dimensional cuts
of $I(1,2,3,4)$ and $I(1,2,4,3)$ are in fact identical: \beq
\label{symm33}
 I(1,2,3,4)|_{s\text{-cut}} \ + \ \ell_1 \leftrightarrow \ell_2 \ = \
 I(1,2,4,3)|_{s\text{-cut}} \ + \ \ell_1 \leftrightarrow \ell_2 \ .
 \eeq
In order to do so,  we use \eqref{iscut} to write
\begin{align}
\label{symm11}
  I(1,2,3,4)|_{s\text{-cut}}\ + \ \ell_1 \leftrightarrow \ell_2
   & = s \langle 4 1 \rangle
  \left(
    \frac{1}{\langle 4 | \ell_1 | 1 \rangle}
    + \frac{1}{\langle 4 | \ell_2 | 1 \rangle}
  \right) \\ \nonumber
  & = s \langle 4 1 \rangle
  \left(
  \frac{\langle 4 | \ell_1 + \ell_2 | 1 \rangle}
  {\langle 4 | \ell_1 | 1 \rangle \langle 4 | \ell_2 | 1 \rangle}
  \right) \\ \nonumber
  & = \frac{s \, \langle 4 1 \rangle \langle 4 | 2 | 1 \rangle}
  {\langle 4 | \ell_1 | 1 \rangle \langle 4 | \ell_2 | 1 \rangle}
  \, ,
\end{align}
where in the last step momentum conservation was used.  Again using
\eqref{iscut} this time for the $s$-cut of $I(1,2,4,3)$ one can write,
\begin{equation}
  I(1,2,4,3)|_{s\text{-cut}}
  = \frac{s \langle 3 1 \rangle}
  {\langle 3  \ell_1 \rangle \langle  \ell_1 1 \rangle} \, ,
\end{equation}
and hence%
\begin{align}
\label{symm12}
  I(1,2,4,3)|_{s\text{-cut}}\ + \ \ell_1 \leftrightarrow \ell_2
  & = s \langle 3 1 \rangle
  \left(
    \frac{\langle 3 | \ell_1 + \ell_2 | 1 \rangle}
    {\langle 3 | \ell_1 | 1 \rangle \langle 3 | \ell_2 | 1 \rangle}
  \right) \\ \nonumber
  & =
    \frac{\langle 3 1 \rangle \langle 3 | 2 | 1 \rangle}
    {\langle 3 | \ell_1 | 1 \rangle \langle 3 | \ell_2 | 1 \rangle} \, .
\end{align}
Next we compare \eqref{symm11} to \eqref{symm12}:
\begin{align}
  \frac{I(1,2,3,4)|_{s\text{-cut}}}
        {I(1,2,4,3)|_{s\text{-cut}}}
        & =
        \frac
            {\langle 4 1 \rangle \langle 4 | 2 | 1 \rangle}
            {\langle 3 1 \rangle \langle 3 | 2 | 1 \rangle}
        \frac
            {\langle 3 | \ell_1 | 1 \rangle \langle 3 | \ell_2 |1 \rangle}
            {\langle 4 | \ell_1 | 1 \rangle  \langle 4 | \ell_2 | 1 \rangle} \\ \nonumber
        & =
        \frac
          {\langle 1 | 4 | 2 \rangle}
          {\langle 1| 3 | 2 \rangle}
        \frac
          {\langle \ell_1 | 3 | \ell_2 \rangle}
          {\langle \ell_1 | 4 |
        \ell_2 \rangle} \\ \nonumber
        & = 1 \, ,
\end{align}
thus proving \eqref{symm33}.
%
% subsection Interchange of $\ell_1 \leftrigh$<++> (end)
% section box-prop (end)

\section{Details on the evaluation of  integrals}
\label{sec:evaluating-integrals}
The integral in our result \eqref{eq:resultsymb} can be reduced to a
set of four scalar, single-scale, master integrals using integration
by parts identities and the \verb FIRE ~package \cite{Smirnov:2008iw}
for \verb Mathematica. In this appendix we present the details of this
reduction as well as the values of these master integrals.
\subsection{Two-loop master integrals in three dimensions}
The master integrals that appear at two loops, in particular in our result
appear in the reduction of our result \eqref{eq:resultsymb}, are given in
 $D=3-2\epsilon$
dimensions by the following expressions:
{
\setlength{\jot}{10pt}
\begin{align}
\text{\bf SUNSET}(q^2)
&
= \begin{minipage}{45px}\sunset\end{minipage}
  =
  -
  \left(\frac{-q^2 }{\mu^2}\right)^{-2\epsilon}
  \frac{\Gamma\left(\frac{1}{2}-\epsilon\right)^3\Gamma\left(2\epsilon\right)}
  {\Gamma\left(\frac{3}{2}-3\epsilon\right)}\label{eq:sunset}\\
\text{\bf TRI}(q^2) &
=
\begin{minipage}{45px}\tri\end{minipage}
  =
  -(-q^2)^{-1}
  \left(\frac{-q^2 }{\mu^2}\right)^{-2\epsilon}
  \frac{\Gamma
    \left(
    \frac{1}{2}-\epsilon\right)^2\Gamma\left(-2\epsilon
    \right)
    \Gamma\left(\frac{3}{2}+\epsilon\right)
  \Gamma\left(2 + 2 \epsilon \right)}
  {\epsilon(1 + 2 \epsilon)^2 \Gamma
    \left( \frac{1} {2} - 3 \epsilon \right)}
    \label{eq:tri}\\
\text{\bf GLASS}(q^2) &
=
\begin{minipage}{70px}\glass\end{minipage}
  =(-q^2)^{-1}
  \left(\frac{-q^2 }{\mu^2}\right)^{-2\epsilon}
  \frac{\Gamma\left(\frac{1}{2}
  - \epsilon\right)^4\Gamma\left(\frac{1}{2}
  +
  \epsilon\right)^2}{\Gamma\left(1-2\epsilon\right)^2}
  \label{eq:glass}\\
  \text{\bf TrianX}(q^2)&\begin{aligned}[t] &
  =
  \begin{minipage}{70px}\centering\trianx\end{minipage}
  =(-q^2)^{-3}
  \left(\frac{-q^2 }{\mu^2}\right)^{-2\epsilon}\,
  e^{-2\gamma_E\epsilon} \biggl[ \frac {4\pi} {\epsilon^2}
    + \frac{\pi (3 + 8\log{2})}{\epsilon}\\
  & - \frac {2\pi} {3}  \left (81 + 4\pi^2
  + 6\log{2}\, ( 4\log{2} -9)\right)
  + \frac {\pi} {6}\biggl(-\pi^2(7 + 40 \log{2})\\
  &
  + 8 \bigl(69 + 6 \log{2} + 2 \log^2{2}(8\log2-27)
  -113\zeta_3\bigr) \biggr) \epsilon + \cO(\epsilon)\biggr]\, ,
\end{aligned}
\label{eq:trianx}
\end{align}
}where we use the conventions of \cite{Czakon:2005rk} for the
integration measure.  The first three of these integrals are planar
and their expressions in all orders in $\epsilon$ can be easily
obtained by first computing their Mellin-Barnes representations most
conveniently using the \verb AMBRE ~package~\cite{Gluza:2010rn} and
then performing the Mellin-Barnes integrations using the MB tools, in
particular \verb MB.m ~\cite{Czakon:2005rk} and \verb barnesroutines.m ~
by David Kosower. The expansion around $\eps=0$ of the {\bf TRI} and {\bf
GLASS} topologies has uniform degree of transcendentality, while this is
not the case for the {\bf SUNSET} and {\bf TrianX} topologies.

\subsection{Reduction to master integrals}
Here we present the reductions of the integral \eqref{eq:intXT} that
appears in our result \eqref{eq:resultsymb} in terms of the master
integrals \eqref{eq:sunset}--\eqref{eq:trianx} of
the previous section:
%{ \setlength{\jot}{10pt}
\beqa
  \label{eq:redxt}
      \text{\bf XT}(q^2)  &\!\!\!\!=\!\!\!\!&
     \frac{7(D-3)(3D-10)(3D-8)}{2(D-4)^2(2D-7)}
      \text{\bf SUNSET}(q^2)\\
  &\!\!\!\!+\!\!\!\!&(-q^2)\frac{5(D-3)(3D-10)}{2(D-4)(2D-7)}
      \text{\bf TRI}(q^2) +(-q^2)^3\frac{D-4}{4(2D-7)}\text{\bf TrianX}(q^2)\,  .
 \nonumber
  \eeqa
%}
Note that the GLASS topology does not appear in $\text{\bf XT}(q^2)$.
Two other integrals we have considered are:
\beqa
  \text{\bf LT}(q^2)
  &  = &
    \frac{8-3D}{D-3}\,
    \text{\bf SUNSET}(q^2)+q^2\left(\text{\bf GLASS}(q^2)-\text{\bf TRI}(q^2)\right)\label{eq:redlt}
    \, , \\
  \text{\bf CT}(q^2)  &
    = &
    \text{\bf FAN}(q^2)
    =
    \left(\frac{1}{4\eps}-\frac{3}{2}\right)\text{\bf SUNSET}(q^2)\, .
\eeqa

%%%%%%%%%%%%%%%%%%%%%%%%%%%%%%%%%%%%%%%%%%%%%%%%%%%%%%%%%%%%%%%%%%%%%%%%%%%%%%%%%%%%%%%%%%%%%%%%%%%%

%\newpage

\end{document}